\documentclass[11pt]{article}

\usepackage[utf8]{inputenc}
\usepackage[T1]{fontenc}
\usepackage{fullpage}
\usepackage{url}
\usepackage{booktabs}
\usepackage{amsfonts}
\usepackage{amsmath}
\usepackage{amssymb}
\usepackage{amsthm}
\usepackage{nicefrac}
\usepackage{microtype}
\usepackage{graphicx}
\usepackage{natbib}
\usepackage{doi}
\usepackage{comment}
\usepackage{xcolor}
\usepackage{hyperref}
\hypersetup{
    colorlinks=true,
    linkcolor=blue,
    citecolor=blue,
    urlcolor=blue,
}
\usepackage{tikz}
\usetikzlibrary{positioning,fit,quotes,arrows}
\definecolor{dkgreen}{rgb}{0,0.6,0}
\definecolor{gray}{rgb}{0.5,0.5,0.5}
\definecolor{mauve}{rgb}{0.58,0,0.82}

\definecolor{color1}{HTML}{53446B}
\definecolor{color2}{HTML}{807DBA}
\definecolor{color3}{HTML}{C0AECE}
\definecolor{color4}{HTML}{CE9292}
\definecolor{color5}{HTML}{DAB6B6}
\definecolor{color6}{HTML}{A40407}
\definecolor{color7}{HTML}{7484A8}
\definecolor{color8}{HTML}{F5BD1C}

\theoremstyle{definition}

\theoremstyle{remark}

\newcommand{\cL}{\mathcal{L}}

\newcommand{\argmax}{\mathop{\mathrm{argmax}}}

\definecolor{DarkGreen}{RGB}{0,100,0}

\definecolor{BrickRed}{RGB}{192, 70, 87}

\definecolor{TealBlue}{RGB}{0, 128, 128}

\theoremstyle{plain}

\theoremstyle{remark}

\title{Reinforcement Learning in the Real World: A Survey of Statistical Challenges and Future Directions}

\author{
  Asim H. Gazi$^{1,3,*}$, 
  Yongyi Guo$^{2,*}$, 
  Daiqi Gao$^{3}$, 
  Ziping Xu$^{4}$, 
  Kelly W. Zhang$^{5}$, 
  Susan A. Murphy$^{1,3}$ \\[1em]
  $^1$Department of Computer Science, Harvard University \\
  $^2$Department of Statistics, University of Wisconsin--Madison \\
  $^3$Department of Statistics, Harvard University \\
  $^4$School of Data Science and Society, University of North Carolina at Chapel Hill \\
  $^5$Department of Mathematics, Imperial College London \\[0.5em]
  \texttt{agazi@g.harvard.edu}, 
  \texttt{guo98@wisc.edu}, 
  \texttt{daiqigao@gmail.com}, \\
  \texttt{zipingxu@unc.edu}, 
  \texttt{kelly.zhang@imperial.ac.uk}, 
  \texttt{samurphy@g.harvard.edu} \\[0.5em]
  $^*$Equal contribution
}

\date{}

\begin{document}

\maketitle

\begin{abstract}
Reinforcement learning (RL) has achieved remarkable success in real-world decision-making across diverse domains, including gaming, robotics, online advertising, public health, and natural language processing. Despite these advances, a substantial gap remains between RL research and its deployment in many practical settings. Two recurring challenges often underlie this gap. First, many settings offer limited opportunity for the agent to interact extensively with the target environment due to practical constraints. Second, many target environments often undergo substantial changes, requiring redesign
%\sam{do we mean redesign RL system OR do we mean use RL to re-optimize the previously learned policy?} 
and redeployment of RL systems (e.g., advancements in science and technology that change the landscape of healthcare delivery). Addressing these challenges and bridging the gap between basic research and application requires theory and methodology that directly inform the design, implementation, and continual improvement of RL systems in real-world settings.

In this paper, we frame the application of RL in practice as a three-component process: (i) online learning and optimization during deployment, (ii) post- or between-deployment offline analyses, and (iii) repeated cycles of deployment and redeployment to continually improve the RL system. We provide a narrative review of recent advances that address
the statistical challenges arising across these three components,
including methods for enhancing sample efficiency during online deployment, maximizing data utility for post- or between-deployment inference, and designing sequences of deployments for continual improvement.
We also outline future research directions in RL that are use-inspired---aiming for impactful application of RL in practice.
\end{abstract}


\section{Introduction}
\label{sec_intro}
Reinforcement learning (RL) has received considerable interest in recent years due to superhuman or state-of-the-art performance in areas such as gaming, autonomous driving, and robotics \citep{Mnih2015, Silver2017, Wurman2022, Tang2025, Kaufmann2023}. One of the factors behind these real-world successes is the proliferation of deep learning techniques that enable flexible learning from big data. Common characteristics of applications where deep RL has success include (1) areas in which the RL algorithm can interact extensively with the target environment or a high-fidelity simulator, and (2) minimal changes in the environment's key variables and dynamics such that re-learning via online RL 
%\sam{redesign of what?--the RL alg or the previously learned policy?} 
and/or  continual redesign of the RL algorithm is often not required  or only infrequently required  for successful decision-making by the algorithm in future deployments. 
For example, a recent high-profile example of RL success in the real world is the use of deep RL to defeat champion human racers in Gran Turismo \citep{Wurman2022}, a highly realistic automobile racing simulator. In this work, the authors train a deep RL agent entirely offline in simulation—using population-based training, rather than pure self-play, to improve robustness to diverse opponent behaviors—and subsequently deploy the learned policy as a fixed controller without online learning or algorithm updates. This success relies in part on the availability of a high-fidelity simulator, which permits extensive interaction with the target environment and large-scale exploration without real-world risks such as automobile crashes. Moreover, although exogenous factors such as opponent behavior may vary across races, the core components of the decision problem (e.g., state representation, action space, reward definition, and governing dynamics) do not change substantially across deployments. As a result, the learned policy does not require continual re-learning or algorithmic modification to remain effective in future use.

{When the two aforementioned characteristics do not apply, RL can be far more challenging in the real world \citep{Dulac-Arnold_realWorldRL_2021}. Specifically, we focus on \textbf{two challenges}:}
\begin{itemize}
    \item \textbf{{Challenge C1:}} {the inability to interact extensively with the target environment or a high-fidelity simulator to obtain a large amount of data, and}
    \item \textbf{{Challenge C2:}} {the presence of significant changes in the environment that require %re-optimizing the decision-making policy by the RL algorithm during the next deployment and/or 
    updating the RL algorithm for use the next deployment.}
\end{itemize}
These challenges are especially salient when RL systems interact with humans. Accordingly, this review focuses on RL systems that interact with humans and aim to influence, support, or respond to potentially changing human behavior. Examples include adaptive interventions in healthcare, intelligent tutoring systems in education, recommender systems and online marketplaces, and human–robot interaction settings in which the robot must learn, support, or respond to human behavior. In such settings, exploration is constrained by the risk of harming a human being, burdening them, or causing them to disengage from interacting with the system. Human needs, their availability to interact, and preferences for interacting with a RL system may also vary across individuals and evolve across deployments.

%An example where RL algorithms interact with humans, have had some recent real-world success, but still require substantial research and development is the use of RL to personalize just-in-time adaptive interventions (JITAIs) \citep{gazikey}. 
For example, the use of RL to personalize just-in-time adaptive interventions (JITAIs) is an application that falls within this scope and highlights both challenges \textbf{C1} and \textbf{C2} \citep{gazikey}.
JITAIs are intervention systems that make sequences of decisions during everyday life to support an individual with their health. RL has gained traction as an online learning and optimization approach for personalizing JITAIs. RL algorithms can be used to select actions about which intervention option to deliver (e.g., whether or not to deliver a push notification-based health intervention to an individual's phone) based on state information sensed and surveyed from the individual (e.g., heart rate or the individual's perceived stress). The RL algorithm then observes how the individual responds and, using this data, learns online to optimize future decisions accordingly. Although RL has shown promise for personalizing JITAIs in recent clinical trials \citep{Aguilera2024, Lauffenburger2024, Trella2025, Ghosh2025, Lee2025_PEARL},
challenges remain that have limited further translation. Related to challenges \textbf{C1} and \textbf{C2} above, every individual and deployment is different, so even if extensive previous data exist, the mismatch between individuals in a new deployment and individuals in a previous deployment may be significant. Society and digital health technology also change rapidly over the span of only a few years, resulting in changing relationships between action, reward, or state variables, as well as changes in the variables themselves \citep{Abernethy2022, Gazi_JITAITwins}. Online learning can help account for these mismatches, but exploration is limited by the number of interactions with an individual---complete disengagement from an intervention can occur due to poor delivery (e.g., removal of the JITAI application) \citep{Nahum-Shani2018}.

\begin{figure}
\includegraphics[width=\linewidth,keepaspectratio]{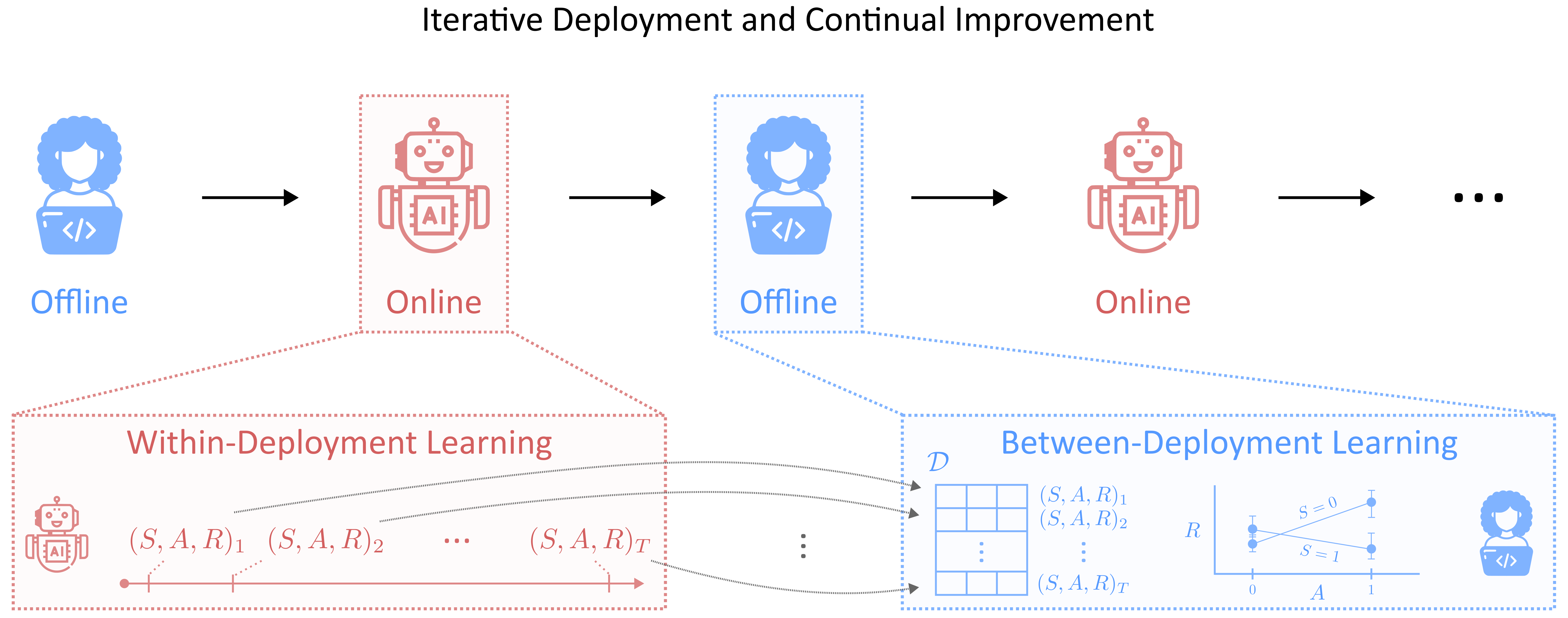}
\caption{Simplified abstraction of reinforcement learning (RL) in practice, framed as a three-component process. The process involves online learning and optimization within each deployment, offline learning and inference between deployments, and a continual deployment-redeployment feedback process for continual improvement. By ``deployment," we refer to the deployment of an RL algorithm, rather than the deployment of an entire system or platform within which an algorithm is deployed. In practice, the online and offline phases can co-occur in time, but this sequential depiction highlights the opportunities for research in offline-to-online learning, online-to-offline learning, and continual improvement in real-world deployments. Also highlighted is the critical role of human oversight in offline analyses and the corresponding revisions needed to update online RL algorithms.}
\label{fig_intro}
\end{figure}

%In this paper we consider settings where RL is poised to make an impact on individuals, such as with  JITAIs and other applications discussed in Sec. \ref{sec_realWorld}. 
For RL to make a substantial impact in the aforementioned real-world applications, challenges  \textbf{C1} and \textbf{C2} must be addressed.  Fig. \ref{fig_intro} provides a forward-looking framework of deploying RL in real-world systems: a three-component process involving online autonomous learning and optimization while a RL algorithm is deployed, offline learning and statistical inference between deployments of RL algorithms, and a continuing deployment-redeployment process for continual improvement over the course of a system's lifetime.
Note that in practice, the online and offline phases may not be as neatly time-aligned as in Figure \ref{fig_intro}. A batch of data obtained via an online RL algorithm's interactions with the environment until a timepoint $t_0$ may be used for offline RL, simultaneously while the online RL algorithm remains in deployment for $t > t_0$. The conclusions from offline learning may then be used to update or deploy a new online RL algorithm at some timepoint $t_1 > t_0$ while the next batch of data from $t_0$ to $t_1$ is considered for offline learning, and so on. Figure \ref{fig_intro} abstracts this into a sequential process for conceptual clarity and to highlight the information transfer from online to offline and back in real-world deployments.
This continuing deployment-redeployment process is related to work on offline-to-online RL \citep{Guo2024_offline2online, Ball2023_O2O, Kim2024_O2O}, where policies learned offline are fine tuned via online RL, but this framework is broader and includes not only offline-to-online RL (i.e., warm starting the next deployment), but also online-to-offline knowledge transfer (i.e., creating generalizable knowledge from each deployment), and the continual offline-to-online-to-offline-... iterative deployment process. %Viewing each deployment as  a sequentially randomized experiment, %(possibly with an adaptive intervention included - see Sec. \ref{sec_realWorld}),
%\sam{I am becoming increasely adverse to use of term "adaptive" --way too much confusion. I prefer to connect to the vast causal literature on sequentially randomized trials (in that area this is sequential randomization within a person).  Adaptive experiments are typically imply bandit environments and further many statisticians will not know what an adaptive intervention is and will confuse this with adaptive clinical trials.}
Statistical methods are critical to enhancing sample efficiency in transitioning from offline to online, as well as ensuring the scientific utility of sequentially randomized data in transitioning from online to offline. 

%The objective of this paper is to provide a survey (i.e., narrative review) of advancements in RL that we find most relevant to future research opportunities poised to make an impact in real-world applications. 
The objective of this paper is to provide a narrative review of methodological developments in RL that we believe are most relevant to addressing challenges \textbf{C1} and \textbf{C2}.
Our goal is not to systematically review or provide a comprehensive survey of RL but rather to focus on research that can inform the design, evaluation, and continual improvement of RL systems that interact with humans and aim to influence, support, or adapt to changing human behavior -- settings in which challenges \textbf{C1} and \textbf{C2} are especially prevalent. Throughout the paper, we emphasize statistical and methodological developments that are use-inspired rather than solely theoretically motivated.
%This survey is centered around the framework shown in Fig. \ref{fig_intro}. 
Multiple surveys and narrative reviews of RL more broadly \citep{Kaelbling1996}, or specific subareas within RL (e.g., deep RL) \citep{Arulkumaran2017}, exist. Recent surveys also include narrative reviews that focus on offline RL \citep{levine2020offline, Prudencio2023}, or exploration and online RL \citep{Ladosz2022}.
In this paper, we do not treat online RL and offline RL as separate, isolated topics. Instead, we organize the review around the three-component framework shown in Fig. \ref{fig_intro}: within-deployment online learning and optimization, between-deployment offline learning and inference, and continual improvement through iterative offline-to-online and back-to-offline cycles.
%However, no reviews to our knowledge highlight the continual offline-to-online \textit{and back to offline} process required to continually improve real-world systems. 
%The review sections of the paper are organized to highlight all three of these components in the framework shown in Fig. \ref{fig_intro}: within-deployment (online, autonomous) learning and optimization, between-deployment (offline) learning and optimization, and continual improvement. %  Throughout we highlight potential approaches to meeting challenges (1) and (2). 
Prior to the review sections, we present basic concepts and notations in Sec.~\ref{sec_notation}, followed by motivating example applications of RL in Sec.~\ref{sec_realWorld} that highlight the three-component process shown in Fig. \ref{fig_intro}. We conclude the paper with a case study in Sec.~\ref{sec:HeartSteps} to elucidate how the challenges and opportunities we present are directly applicable to a concrete real-world example.

% \sam{comments by Ziping: Regarding section 1, I slightly worry that we will upset people when we say that of their environment’s key variables and dynamics are stationary such that continual redesign and re-learning via online RL is not required. As we all know a lot of work on domain adaptations, transfer learning, meta learning has been done to generalize a learned agent to new tasks or from sim to real. 

% I feel the better way to say is that the fact we don’t have a high-fidelity simulator or large scale dataset makes non-stationarity issue a lot more challenging. 
% }

%\sam{The intro, section 1. is sets the stage with the three component process of Figure 1 and the two challenges.  To tie the paper together we should repeatedly return to this. }

\section{Key Concepts and Notation}
\label{sec_notation}
%In this section, we provide an overview of key definitions in reinforcement learning (RL)  \citep{sutton1998reinforcement}. 

In this section, we review key definitions in RL that are essential for our discussion. For more comprehensive overviews of RL concepts and notation, see \citet{kaelbling1996reinforcement, sutton1998reinforcement, puterman2014markov, lattimore2020bandit, szepesvari2022algorithms}. 

%\asim{do we need to define agent or help people understand the difference between a RL algorithm and a specific RL agent as an instantiation of the algorithm? Or will people know?} 

In reinforcement learning (RL), an agent sequentially makes decisions (i.e., selects actions) over time by interacting with an environment that generates states and rewards in response to the agent’s actions.\footnote{Here, an RL agent can be understood as a concrete instantiation of an RL algorithm---i.e., the algorithm with specific parameters and randomness running in the environment.} Specifically, at each decision time $t$, an \textit{agent} selects an \textit{action}, $A_t$, based on its observation of the environment's \textit{state}, $S_t$, and then receives a \textit{reward}, $R_t \in \mathbb{R}$, and the environment transitions to the next state $S_{t+1}$.  Interactions with the environment create data tuples, $(S_t, A_t, R_t)$, 
indexed by decision time $t$. In practice, depending on the application, the environment may correspond to a single individual, a cohort of individuals, or a non-human system such as a robotic platform. The decision times may be very frequent such as every millisecond or every minute, or twice daily and so on, depending on the problem. The state $S_t$ consists of variables that capture the current status of each unit. For instance, the state may encode a patient's health indicators at each decision time. The action $A_t$ represents the decision option that was selected by a decision-making agent, such as whether or not to deliver a treatment. Here we assume that the set of actions is finite. The reward $R_t$ quantifies the optimization goal for the agent, such as the value of a biomarker that should be maximized. We assume that the reward is bounded and is coded such that high values are preferred. %The \textit{environment} is the setting in which the RL agent makes decisions and receives feedback.

The \textit{history} at time $t$ is defined as
$$
\mathcal H_t = (S_1, A_1, R_1, \ldots, S_t, A_t, R_t),
$$  
which records the sequence of interactions between the agent and the environment. In general, the environment evolves probabilistically such that each state and reward may depend on the entire past history: at time $t$, the state is generated as $S_t \sim P_{S,t}(\cdot \mid \mathcal H_{t-1})$; then given an action, $A_t$, the reward is generated as $R_t\sim P_{R,t}(\cdot \mid \mathcal H_{t-1}, S_t, A_t)$. The agent selects actions according to a \textit{policy}, denoted by $\pi$, which specifies, for each time $t$, a conditional distribution over actions given the past history and current state.
Specifically, we write $\pi = \{\pi_t\}_{t\ge 1}$, where $A_t\sim \pi_t(\cdot \mid \mathcal H_{t-1}, S_t)$. As discussed below, online RL algorithms determine this policy, which is often dependent on the history and  nonstationary.
%The agent makes each decision $A_t$ based on a \textit{policy}, denoted by $\pi$, which consists of the conditional distributions of each action $A_t$ given the history prior to $A_t$: $\pi = \{\pi_t\}_{t\ge 1}$ where $A_t\sim \pi_t(\cdot \mid \mathcal H_{t-1}, S_t)$. As discussed below online RL algorithms determine this policy which is often dependent on the history and  non-stationary.
%This fully history-dependent environment is highly general and can represent complex sequential decision-making applications. %Because this environment is often too general to yield tractable algorithms,
%the state $S_t$ is drawn conditional on $\mathcal H_{t-1}$; the action $A_t$ is selected from a \emph{policy} $\pi = \{\pi_t\}_{t=1}^T$, with $\pi_t$ specifying a distribution over actions given $(\mathcal H_{t-1}, S_t)$; The reward $R_t$ is subsequently generated based on $(\mathcal H_{t-1}, S_t, A_t)$. The interaction repeats over a horizon $T$ (finite or infinite), producing the full trajectory.
Together, the environment dynamics and the  policy $\pi$ induce a joint distribution over the history $\mathcal H_{t}$. The probability density for $\mathcal H_{t}$ can be written as
\begin{align}\label{eq::history-density-general}
&p_0(s_1)\pi_1(a_1 \mid s_1)p_{R,1}(r_1\mid s_1, a_1)\cdot\nonumber\\
&\prod_{\tau=2}^t p_{S,\tau}(s_{\tau} \mid  h_{\tau-1})\pi_\tau(a_\tau \mid s_\tau,  h_{\tau-1}) p_{R,\tau}(r_\tau\mid  h_{\tau-1}, s_\tau, a_\tau),
\end{align}
where the lower case $p$ indicate an associated probability density and $p_0$ is the probability density for the initial state, $S_1$.
In the following, we will index this distribution by the policy $\pi = \{\pi_t\}_{t\ge 1}$ and thus will use the notation $\mathbb{E}_\pi$ to denote expectation with respect to this distribution.

\subsection{Common models of RL environments}
\label{common_environments}

\begin{figure}[t]
    \centering
    \includegraphics[width=0.9\linewidth]{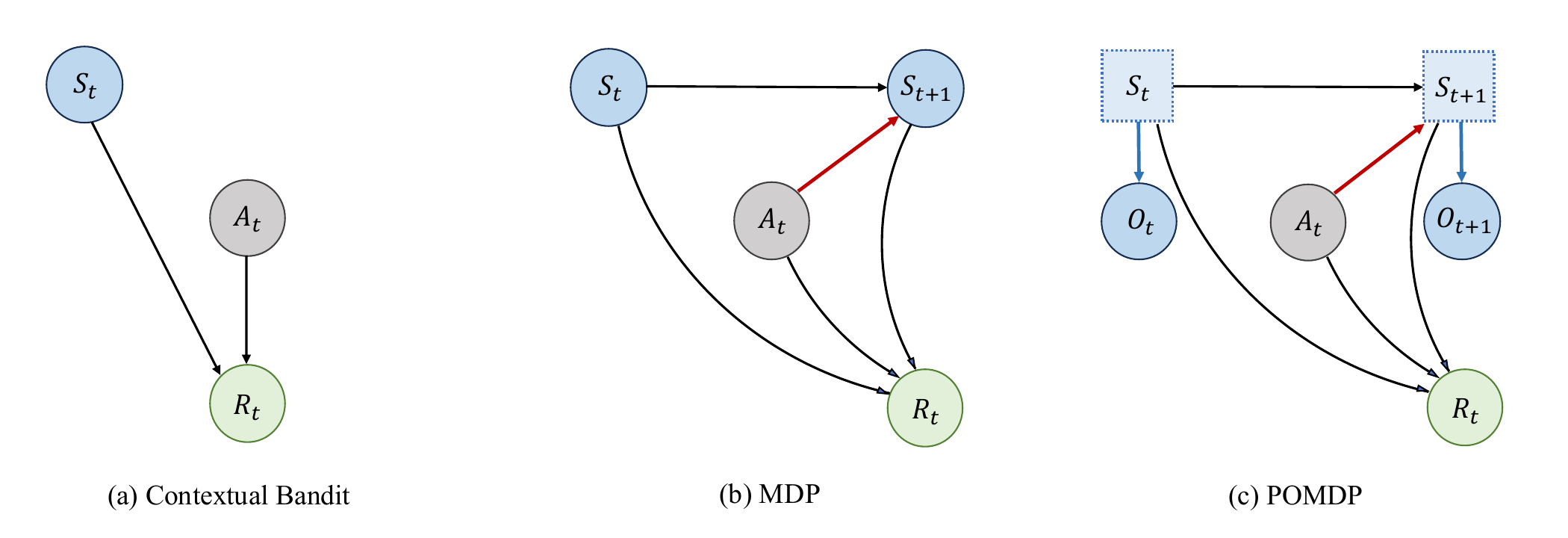}
    \vspace{-2mm}
    \caption{Illustrative causal diagrams of sequential decision-making environments: (a) contextual bandit, (b) Markov decision process (MDP), and (c) partially observable MDP (POMDP). Note that the action $A_t$ is selected by the agent and may depend on the entire history of past observations. 
    In the MDP diagram (b), the red arrow highlights the effect of the action $A_t$ on the next state $S_{t+1}$, which is absent in the contextual bandit setting (a). In the POMDP diagram (c), the blue arrows indicate the additional observation process, whereby the agent observes $O_t$ rather than the latent state $S_t$; dotted squares indicate unobserved (latent) states.
    %The blue arrows in (c) capture additional relationships in the POMDP environment not present in (b). The dotted squares in (c) indicate that the state is not observed by the agent. 
    Other examples of causal diagrams for the environment  include those  used in causal RL; see, e.g., \citet{bareinboim2021introduction, ge2025review}. %\sam{In classical bandits there is no arrow from one state to the next. } 
    \vspace{-3mm}}
    \label{fig:rl_dag}
\end{figure}

While the formulation above allows for fully history-dependent environments, most RL methods are developed and analyzed under additional structural assumptions on the environment. In this section, we discuss common models of RL environments. Quite confusingly, these environments are often paired with classes of online RL algorithms referred to by the {\it same names}. 

\textbf{Markov Decision Processes  (MDPs)}. MDP environments assume  the environment’s state dynamics possess the \textit{Markov property}. This implies that the joint distribution of the reward and next state is conditionally independent of the history, given the current state–action pair. Unless noted otherwise, here MDP state dynamics  are assumed  stationary (time-homogeneous or time-invariant) and, for simplicity, the reward $R_t$ is a known, possibly stochastic, function of $(S_t,A_t, S_{t+1})$. Given a policy, $\pi = \{\pi_t\}_{t\ge 1}$ which is history-dependent and nonstationary as defined above, the density for $(\mathcal H_{t}, S_{t+1})$ can be written as\footnote{This decomposition is reorganized from \eqref{eq::history-density-general} to make explicit the causal ordering among state, action, and reward variables in an MDP, as depicted in Fig.~\ref{fig:rl_dag}(b). It is a special case of \eqref{eq::history-density-general} obtained by imposing the Markov assumption on the environment dynamics. In the general longitudinal formulation (\ref{eq::history-density-general}), the reward density at time $\tau$ is $p_{R,\tau}(r_\tau\mid  h_{\tau-1}, s_\tau, a_\tau)$ and does not explicitly condition on $s_{\tau+1}$. This is without loss of generality: the dependence is simply not made explicit because the reward and next-state distributions are specified separately in the product. Indeed, (\ref{eq::history-density-general}) places no restriction on the joint dependence structure of $(r_\tau, s_{\tau+1})$ on $(h_{\tau-1}, s_\tau, a_\tau)$.}
%\begin{align*}
%p_0(s_1)\pi_1(a_1 \mid s_1)p_{R,1}(r_1\mid s_1, a_1)\prod_{\tau=2}^t p_{S,\tau}(s_{\tau} \mid  s_{\tau-1}, a_\tau, r_\tau)\pi_\tau(a_\tau \mid s_\tau,  h_{\tau-1}) p_{R,\tau}(r_\tau\mid  s_\tau, a_\tau),
%\end{align*}
%\begin{eqnarray*}
%p_0(s_1)\pi_1(a_1 \mid s_1)p_{R}(r_1\mid   s_1, a_1)\prod_{\tau=2}^t p_{S}(s_{\tau} \mid  s_{\tau-1}, a_{\tau-1}, r_{\tau-1})\pi_\tau(a_\tau \mid s_\tau,  h_{\tau-1}) p_{R}(r_\tau\mid   s_\tau, a_\tau).
%\end{eqnarray*}
\begin{eqnarray*}
p_0(s_1)\prod_{\tau=1}^t \pi_\tau(a_\tau \mid s_\tau,  h_{\tau-1}) p_{S}(s_{\tau+1} \mid  s_{\tau}, a_{\tau}) p_{R}(r_\tau\mid s_\tau, a_\tau, s_{\tau + 1}).
\end{eqnarray*}
The density of $\mathcal H_t$ can be obtained by integrating out $S_{t+1}$. The relationships between states, actions, and rewards in an MDP are illustrated in the causal directed acyclic graph (DAG) in Fig. \ref{fig:rl_dag}(b). Here, a causal DAG is an acyclic graph that uses vertices and directed edges to represent variables and their causal relationships \citep{pearl2009causality}. It visually helps researchers understand the causal relations between sets of variables through causal reasoning and causal discovery.

%\ziping{In Fig 2, there is no arrow from $s_{t+1}$ to $r_t$. Better to keep it consistent here. In page 4, there is also no $s_{t+1}$ in the reward generation. } 
%\sam{Asim, note (1) the density above contains the conditional density of  $r_t$ given $(s_t, a_t, s_{t+1})$; (2) in the RL field the reward, $r_t$ is often assumed to be a known function, $f$ of $(s_t, a_t, s_{t+1})$, $r_t=f(s_t, a_t, s_{t+1})$ or $r_t=f(s_t, a_t, s_{t+1}) + \epsilon_t$ where $\epsilon_t$ is exogeneous --thus the figures are confusing.. }
%:$$ P(S_{t+1}, R_t|\mathcal H_{t-1}, S_t, A_t) = P(S_{t+1}, R_t|S_t, A_t).$$
%\YG{Do not need this paragraph} Formally, an MDP is specified by the tuple $(\mathcal S, \mathcal A, P, R)$, where where $\mathcal S$ and $\mathcal A$ are the state and action spaces defined above, and 
%\begin{itemize}
%    %\item $\mathcal S$: the set of states, \ziping{These notations are repetitive.}
%    %\item $\mathcal A$: the set of actions,
%    \item $P(\cdot|s,a)$ is the transition kernel specifying the distribution of the next state given current state–action pair $(s, a)$,
%    \item $R(\cdot|s,a)$ is the distribution of rewards under state–action pair $(s, a)$.
%\end{itemize}

Next, we consider three commonly studied special cases of Markov decision processes where an optimal policy might be defined through the maximization of expected cumulative rewards.

\textbf{Finite Horizon MDP.} Finite-horizon MDPs run in episodes. Each episode begins at an a initial state, sampled from $p_0$, proceeds for $H$ steps under the MDP dynamics, and then terminates. An optimal policy is therefore defined over $H$ number of decision times, and $H$ is called the horizon.
%The optimal policy is defined over  a finite number of decision times $H$ (horizon is $H$). 
An example is in health applications, the intervention for each patient can be designed to last for a fixed period of time (i.e., the intervention is at most $H$ decision times). 
%Finite-horizon MDPs run in episodes. 
%Each episode begins at an i.i.d sampled  initial state, proceeds for $H$ steps under the MDP dynamics, and then terminates.
For a given policy $\pi$, the \textit{state-value function} and \textit{state-action-value function}   are defined as
\[
V^\pi(s) = \mathbb{E}_\pi \!\left[\sum_{t=1}^H R_t \,\middle|\, S_1 = s \right], 
\quad
Q^\pi(s,a) = \mathbb{E}_\pi \!\left[\sum_{t=1}^H R_t \,\middle|\, S_1 = s, A_1 = a \right].
\]
These capture the expected cumulative reward starting from state $s$, with or without conditioning on a specific initial action.
%The RL agent's performance is often measured by regret, which quantifies the loss incurred by following the agent's policy instead of the \textit{optimal policy} (unknown to the agent): for any number of episodes, $T$,  the (expected) regret of the agent's policy $\pi$ is given by 
%\[\text{Regret}(T; \pi) = \mathbb{E}\left[\sum_{k=1}^T V^{*}(S_{1,k}) - V^{\pi}(S_{1,k})\right],\]
%where $S_{1,k}$ is the initial state of episode $k$, and the outer expectation is taken with respect to the initial state distribution.  
An optimal policy is defined as $\pi^* = \arg\sup_{\pi}V^{\pi}(s)$; in finite-horizon
MDPs an optimal policy is  generally nonstationary in $t$ and history dependent (\cite{puterman2014markov}, Chap. 4). 

 %Here, $V^*(s) = \sup_{\pi}V^{\pi}(s)$ is the optimal state value, attained by some optimal policy $\pi^*$

\textbf{Finite Horizon MDP with $H=1$: Contextual Bandits.}\label{sec::setting-bandit}
MDP environments with finite horizon, $H=1$, are so common that they have their own name, ``contextual bandits," and  separate notation. 
%An episode is only $(S, A, R)$.  
The relationships between states, actions, and rewards over time in a contextual bandit are illustrated in the causal diagram in Fig. \ref{fig:rl_dag}(a).
In this literature the state, $S$, is often called the context or side information, and actions are commonly called arms.
%Further  $t$ is often used to index the episode and episodes are referred to as decision times.  Here we will use $k$ instead of $t$ and we will use the term, episode, for consistency with the above description of the general $H$ finite horizon environment.
For a given policy, $\pi$, the value function ($H=1$) becomes 
$V^\pi(s) = \mathbb{E}_\pi \!\left[ R \,\middle|\, S = s \right]$ and the action-value function $
Q(s, a) = \mathbb{E} \!\left[ R \,\middle|\, S=s, A = a \right]$. In the contextual bandit setting $Q(s, a)$ is often called the reward function. 
%As before, $V^*(s) = \sup_{\pi}V^{\pi}(s)$. 
Further, 
%$V^*(s) =\max_\pi \mathbb E_\pi\left[R\,\middle|\, S = s \right]$ and 
the optimal policy is given by $\pi^*\in\argmax_\pi \mathbb E_\pi\left[R\,\middle|\, S = s \right]$.

\textbf{Infinite-horizon discounted MDP.} Here, the optimal policy is defined by  a "softened" version of the finite horizon MDP. By softened version, we mean that  instead of terminating an episode with probability $0$ when $t<H$ and terminating  with probability $1$ when $t=H$, at each time $t$ the decision making process, independent of the MDP, proceeds to the next decision time with probability $\gamma\in[0,1)$ given that the process has not yet terminated (\cite{puterman2014markov}, Chap. 5). 
%\sam{susan disagrees with following...!} \ziping{There is no real termination in infinite-horizon discounted MDPs. It is about the optimization goal. I worry the current writing will confuse readers. In Puterman: Discounting arises naturally in applications in which we account for the time value of the rewards. The discount factor $\gamma$ measures the present value of one unit of currency received one period in the future.} 
This geometric-termination view makes the discounted problem a soft approximation to a finite-horizon problem with an effective horizon $H=\frac{1}{1-\gamma}$. The corresponding value functions are 
\[
V^\pi(s)=\mathbb{E}_\pi\Bigg[\sum_{t=1}^\infty \gamma^{t-1} R_t \,\Big|\, S_1=s\Bigg],
\quad
Q^\pi(s,a)=\mathbb{E}_\pi\!\Bigg[\sum_{t=1}^\infty \gamma^{t-1} R_t \,\Big|\, S_1=s,\;A_1=a\Bigg].
\]
The optimal policy is again defined by 
$\pi^* = \arg\sup_{\pi}V^{\pi}(s)$. In discounted MDPs, an optimal policy can be chosen to be Markovian (depends only on the current state) and stationary (time-invariant) (\cite{puterman2014markov}, Chap. 5). This is one way the "softened" termination---via discounting---yields a less complex objective than the finite-horizon case.

\textbf{Extensions and generality.} One way in which data observed from the environment will appear to be history dependent and nonstationary (even if the underlying state dynamics are Markovian and stationary) is if the environment follows a \emph{partially observable MDP} (POMDP), shown in Fig. \ref{fig:rl_dag}(c).  Here the assumption that the state is fully observed  does not hold. Instead, the environment evolves through \textit{latent} states $S_t$ that follow the same state dynamics and reward generation as in a MDP. However,  $S_t$ is not observable; rather, at each time  only  $O_t$ can be observed. $O_t$ provides only partial or noisy information about the underlying state, $S_t$. %Actions are then chosen based on the history of past rewards, actions, and observations, rather than true states. 

%In a POMDP, the agent instead receives noisy or incomplete observations and must base decisions on these observations.

Note that one can always define the state at time $t$ to be the entire time $t$ history. This would always result in Markovian state dynamics. While such a state representation may be prohibitively large in practice, this observation underscores the generality of MDPs and provides practical insights into state construction when modeling the environment as a MDP.
Beyond this general definition, if a known causal DAG is available to describe the causal relations among all variables in the history, it is possible to identify a lower-dimensional state that retains all necessary information \citep{zhang2020designing,gao2025harnessing} (see details in Sec. ~\ref{sec:bias-variance}).

%\YG{If one has full knowledge of the longitudinal environment, it is always possible to define the state as the entire history so that the Markov property holds. In practice, however, such a state representation can be prohibitively large, but this observation highlights that the MDP formalism is quite general.}

%\sam{ I commented out some text; see .tex file.   We need to clarify exposition.  Below $H$ will denote the horizon.   Yet we used $T$ here and called $T$ the horizon.   The problem is that $T$ often denotes the learning horizon (number of "rounds"), but $H$ denotes the number of decision times.  Sometimes $T=H$ but not always.  For example in episodic RL, $T$ is the number of episodes, whereas $H$ is the duration of each episode.}

% \sam{do you think we should include environments assumed by people in causal RL?  this is a setting in which one uses a causal diagram to describe how these distributions, $P_{S,t}(\cdot \mid \mathcal H_{t-1})$ and $ P_{R,t}(\cdot \mid \mathcal H_{t-1}, S_t, A_t)$ depend on prior state and history. We need Daiqi help here!} \daiqi{Which environments are you referring to? History dependent environments?} \sam{I mean potentially history dependent and then you use the DAG to construct a markovian state.  So just a sentence with reference to that literature in which they use a causal dag to specify the state.  would some of ambuj tewari's work be relevant?  not sure...}

Taken together, the more general history-dependent framework, MDPs, and contextual bandits are the most commonly assumed environments in sequential decision making under uncertainty.  %For common RL algorithms that assume these either the MDP and/or contextual bandit environments, see Sec. \ref{sec_within_trial}.

\subsection{The agent}
Given the environment dynamics described above, we now turn to the \textit{agent}. Recall that an RL agent is a concrete instantiation of an RL algorithm---i.e., an algorithm with specific parameters and internal randomness---that interacts with the environment by selecting actions according to a policy $\pi = \{\pi_t\}_{t\geq 1}$. An online RL algorithm, denoted as $\mathcal{L}$, is an algorithm designed to specify how the policy $\{\pi_t\}_{t\geq 1}$ is updated over time as data are collected. 

The objective of an online RL algorithm is typically to induce a policy that performs well with respect to a given criterion, such as maximizing expected cumulative rewards over the course of a deployment. A central challenge in RL is that the environment dynamics are not fully known a priori; for example, the transition dynamics and/or value functions may be modeled parametrically, semiparametrically, or nonparametrically with unknown components. Consequently, online RL algorithms must learn about the environment from streaming data observed during deployment while simultaneously making decisions that influence the environment. More specifically, at each time $t$, an online RL algorithm uses the history to ``learn'' parameters of the environment, generates a policy $\pi_t$ to ``make decisions", and then executes these decisions in the environment by assigning $A_t$ according to $\pi_t$. This learning-while-acting requirement gives rise to the \emph{exploration–exploitation tradeoff}: the algorithm must balance \emph{exploration}, selecting actions whose rewards are uncertain and possibly suboptimal in order to gain information about the environment, with \emph{exploitation}, choosing actions that appear best given current knowledge, even though that knowledge may be incomplete. 

A common way to evaluate the performance of an online RL algorithm $\mathcal L$ is through the notion of \emph{regret}. While the precise definition of regret depends on the setting and environment, it generally measures the difference in expected cumulative reward between an optimal policy (within a specified policy class) and the policy induced by $\mathcal L$. To make this concrete, consider the contextual bandit setting. Recall that at each time $t$, the agent observes the state $S_t$, selects an action $A_t\sim \pi_t(\cdot\mid \mathcal H_{t-1}, S_t)$, and receives a reward $R_t\sim p_{R}(\cdot|S_t, A_t)$. Let $\pi^*$ be the optimal policy within a certain class. Then the regret of an online learning algorithm $\cL$ up to time $T$ is 
$$
\mathbb E_{\pi^*}\left[\sum_{t=1}^TR_t\right] - \mathbb E_{\cL}\left[\sum_{t=1}^TR_t\right],
$$
where $\mathbb E_{\cL}$ denotes the expectation with respect to the randomness induced jointly by the environment and by the (possibly randomized) policy $\pi$ generated by $\cL$.

See Sec. \ref{sec_within_trial} for more information concerning online RL algorithms.

\section{Motivating Examples}
% \section{Example Applications of RL in Interventions for Humans}
\label{sec_realWorld}
In this section, we summarize two classes of real-world applications of online RL involving interventions for humans: the more established adaptive experiments and the more nascent use of RL to optimize adaptive intervention systems. These example use cases will be used to motivate and ground the remaining discussion. 
The brief subsection on adaptive experiments is included to review the application setting where much of the statistical literature on RL in interventions for humans has focused on to date.
The more extensive subsection on adaptive interventions is included to highlight the more nascent application setting with important differences ripe for future research. 
A common theme across both use cases is that the RL algorithms intervene on people's lives in some way. This often leads to the challenges presented in Sec. \ref{sec_intro}: \textbf{(C1)} extensive exploration data required for the RL system to learn solely online (i.e., within-deployment) is often unavailable, and \textbf{(C2)} significant changes in the deployment environment---such as societal shifts or evolving characteristics of the target population---not only necessitate learning online (i.e., optimizing the policy) to account for shifts between previously collected data and the present day application, but also require human-in-the-loop redesign (i.e., adding/removing actions, data sources), offline learning (i.e., between-deployment), and redeployment to continually improve the online RL algorithm and overall system to mirror changes in the real world (i.e., iterative deployments). %\sam{I think we are talking about domain shift and distribution shift.  We use these phrases in section 5.2}
Note that the challenges and themes we highlight in this section apply broadly to RL application areas involving intervening upon humans (e.g., recommender systems, human-robot interaction, etc.) and are not restricted to the specific examples we describe here.

\subsection{Adaptive Experiments} \label{subsec:adaptiveExp}
% \sam{lets make this first section about adaptive experimental design.  Some of this occurs in education as well as health. Give very very brief examples of both.}   
Adaptive experiments are a mature application of RL. Unlike classical randomized experiments, where units of an experiment (e.g., participants) are assigned to experimental arms using fixed assignment probabilities (e.g., 50–50 in a two-arm trial), adaptive experiments use online RL to adapt these probabilities over time based on data from prior participants. Note that the adaptivity here is part of the experimental design, in contrast with adaptive interventions described in the next subsection, wherein the adaptivity is part of the intervention design. Concrete examples of adaptive experimental design include the assignment of educational interventions to students \citep{Rafferty2019, Reza2021}, as well as the adaptive assignment of participants to experimental arms in a clinical trial \citep{Newby2025, Bhatt2016, Pallmann2018}. In these settings, online RL is used to adapt arm assignment probabilities to favor positive outcomes observed so far.
%Thus, adaptive experiments are often a preferred approach when investigating one or more improvements to a system or intervention that is already in use.
%above statement is false in adaptive clinical trials...
% Adaptive experiments underlie frameworks such as the MOOClet framework \citep{Reza2021}, which aims to augment massive open online courses (MOOCs) with adaptive experimentation tools. 
% Other examples include adaptively assigning participants to experimental arms in a clinical trial for healthcare interventions \citep{Newby2025, Bhatt2016, Pallmann2018}. 

In adaptive experiments, the environment is often assumed to have a finite horizon equal to one (i.e., contextual bandit environment) (see Sec. \ref{sec_notation}), particularly when each assignment decision (i.e., which experimental arm) concerns a different individual independently sampled from a target subpopulation. The actions correspond to the possible experimental arms that each individual can be assigned to, each context vector represents a specific individual's characteristics, and the reward is a function of that individual's outcomes.

\subsection{Adaptive Interventions} 
\label{sec:adaptInterv}
% \sam{this section is about the use of RL as a component in an intervention system. Examples abound in both education and in digital health.  There is a vast number of papers on "intelligent tutoring systems"--many use RL to make sequential decisions on each student.  I put two example papers in the folder papers in this overleaf file.   }

A more nascent application of online RL is in the optimization of adaptive interventions. Adaptive interventions automatically decide at each time point, in a sequence of time points, how to tailor interventions (e.g., healthcare support, educational content, etc.) for a particular individual or group of individuals based on policies that map state variables (e.g., location, cognitive load) to intervention options (e.g., which intervention to provide). 
\textit{The use of RL to optimize an adaptive intervention fundamentally differs from the use of RL in adaptive experiments.} In an adaptive intervention, RL is part of the intervention system, as opposed to part of the design of an experiment. Importantly, adaptive interventions involve sequential decision-making on each individual or group of individuals (i.e., sequences of decisions per individual), unlike adaptive experiments that generally involve a sequence of individuals, with a single decision per individual. Note that since adaptive interventions involve sequences of decisions for each individual, the environment (i.e., the distribution of the individual's longitudinal outcomes) is unlikely to be a contextual bandit environment. %Actions represent intervention options, rewards capture outcomes of interest, and the state vector includes time-varying information on the recipient of the adaptive intervention.

The use of online RL to optimize adaptive intervention policies is motivated by the fact that fixed policies derived from prior intervention data are often suboptimal for new recipients of the adaptive intervention due to societal changes, inclusion of new intervention components in the adaptive intervention, and so on, leading to challenge \textbf{C2}.
Online RL optimizes policies to the new individuals as they experience the adaptive intervention. Adaptive interventions in healthcare include just-in-time adaptive interventions (JITAIs), which automate the delivery of behavioral and/or biological treatment (i.e., select treatment actions) \citep{Nahum-Shani2018, Deliu2024, gazikey} via smart devices (smart wearables, smartphones, smart speakers, etc.). A concrete example is  the ``HeartSteps" JITAI designed to promote physical activity. The HeartSteps JITAI, described further in the case study in Sec. \ref{sec:HeartSteps}, has been repeatedly updated and deployed: the first deployment targeted sedentary adults in the Midwest US, the second and third deployments targeted adults with Stage 1 hypertension in the Northwest US and the fourth deployment targeted adults with obesity in Southwest US, respectively \citep{Klasnja2019, liao2020personalized, de2025dynamic}. In education, adaptive interventions include intelligent tutoring systems (ITSs). ITSs adapt instructional content, feedback, or question difficulty based on a student's knowledge states and engagement \citep{Riedmann2025}. As with JITAIs, ITSs must continually improve and account for new populations, curricula, and observations (e.g., clickstream data, eye tracking, or biosignals).

% \sam{ Here I suggest that we focus on challenges  (in a parallel way to how we did this in adaptive experimentation paragraph above).
% Put current approaches to deal  with challenges later. }

% \sam{Following text on approaches to deal with challenges goes elsewhere? The following overlaps a great deal with the discussion of bias -variance tradeoff. Some of this belongs here as  it discusses challenges but much could go  elsewhere like  maybe where we discuss bias vs variance tradeoff?}

%\sam{go over following and explicitly connect to the two challenges in intro-- also connect **explicitly** to Fig 1.} 
Several technical challenges complicate the design of RL algorithms for adaptive interventions. A primary challenge that corresponds to challenge \textbf{C1} is the scarcity of data available from the individual experiencing the adaptive intervention. Although substantial data may exist from prior recipients of similar interventions, new recipients often differ, and broader societal shifts or technological advances can alter the underlying environment that needs to be approximated by the RL algorithm's models. Since data collected online for any given individual is typically scarce, sample efficient methods are needed for online RL. One common approach to accelerating online learning is to pool data across multiple individuals when making decisions for each individual, thereby trading bias for variance in learning (See Sec. \ref{sec:bias-variance}).  
However, such pooling induces statistical dependence across individuals’ data, which complicates statistical inference and offline learning (See Sec. \ref{sec:inference-pooling}). 
%One approach to improving the amount of data and hopefully  accelerate online learning is for the RL algorithm to leverage, that is, pool data from multiple individuals to make decisions on each individual.  The idea is to tradeoff bias with  variance in learning . 

The second challenge, corresponding to challenge \textbf{C2}, concerns the effective redesign  of RL algorithms across iterative deployments due to changes in the underlying environment. In adaptive health interventions, this challenge is compounded by the high-stakes nature of healthcare deployments. Since the online RL algorithm is part of the intervention system, all components---including the learning algorithm, hyperparameters, and policy optimization---often must be prespecified, especially in clinical trial settings, and cannot be modified within a deployment without invalidating results. This underscores the need for careful thought on how best to tradeoff bias in learning with autonomy and stability of the online RL algorithm.

%“Pooling” data across recipients can accelerate learning, as concurrent participants generate more data collectively.

\subsection{Summary}
The motivating applications reviewed in this section illustrate both the promise and challenges of deploying RL in real-world systems that interact with humans. Environments evolve, human beings can be directly affected by decisions, and data may be scarce. These real-world characteristics again highlight the two challenges we have emphasized throughout the paper thus far: \textbf{(C1)} extensive exploration required to learn solely online may not be available, and \textbf{(C2)} the presence of significant changes (e.g., new interventions in a system) may require either re-optimizing the policy of the RL algorithm or redesigning the RL algorithm itself. To make RL broadly useful in such settings, methodological advances are needed that go beyond minimizing regret in online learning or achieving exact inference offline. Future work must also address offline-to-online sample efficiency, valid and reliable online-to-offline inference, and continual improvement.
%\sam{explicitly bring in the two parts of challenge (2) here... } 

\section{Within-deployment learning and optimization}
\label{sec_within_trial}
%\sam{ should all following sections focus on adaptive intervention design (as opposed to adaptive experimental design).  I think so.  I think we implicitly doing this.  If you agree then we need to explicitly tell reader.} %\YG{Susan: I do not think Section 4 should be restricted to adaptive interventions. The material here focuses on general online RL strategies for within-deployment learning—across both bandit and MDP settings—which are relevant to both adaptive experiments and adaptive interventions. I tried to make some adjustments in writing below.}

In this section, we review and discuss online RL for within-deployment learning and optimization. As noted in Sec.~\ref{sec_notation}, a common performance metric is regret, whose minimization gives rise to the exploration–exploitation tradeoff. To balance this tradeoff, online RL algorithms operate using two interconnected components: \emph{a learning} component, which learns about the environment, and \emph{a decision-making} component, which takes the learned knowledge as input and selects actions to address the exploration–exploitation tradeoff \citep{sutton2018reinforcement}. This decomposition into learning and decision-making guides our discussion below. %\ziping{Is there a paper we can cite about this view point?}\YG{I cannot find a very good reference...}

For the learning component, RL algorithms can be broadly categorized as \emph{model-based} or \emph{model-free}. 
Model-based algorithms learn an explicit (parametric, semiparametric, or nonparametric) statistical model of the environment. In contextual bandit settings, this typically corresponds to specifying and estimating a model for the conditional reward distribution given context and action, while in MDP settings it involves learning the transition dynamics and reward function. Given a learned model, subsequent action selection or planning is handled by a separate decision-making component; in bandits, this often reduces to selecting the action with the highest predicted reward (possibly incorporating optimism or posterior sampling introduced below), whereas in MDPs this may involve classical planning methods such as dynamic programming (Chap.~4 in \citet{sutton2018reinforcement}). For a recent survey of model-based RL, see \cite{m2023model}.
%learn the environment’s transition function (e.g., estimate parameters in a parametric, semi-parametric or non-parametric statistical model for the transition function). Given the transition function, classical methods in the MDP literature such as dynamic programming can be used to construct an optimal policy (see Chap. 4 in \citet{sutton2018reinforcement});
%; a classic example is Dyna-Q \citep{sutton1990integrated}
%for a more recent survey see \cite{m2023model}. 
%, which integrates direct value learning with model-based planning through simulated experience. 
%More recent theoretically grounded methods plan optimistically within confidence sets of the estimated model \citep{kearns2002near, auer2008near}. 
Here we focus on \emph{model-free} RL algorithms, which do not attempt to learn an explicit model of the environment, such as transition dynamics in MDPs or a parametric reward model in bandit settings. 
Instead, these methods learn value functions or policy parameters directly to support policy improvement.
%Here we focus on \emph{model-free} RL algorithms.  These algorithms do not attempt to model the transition function. Rather model-free algorithms use models of marginal quantities such as a state or action-state value function. Online model-free RL usually includes two components. The first component focuses on learning parameters in the marginal model.
%\sam{would be better for reader if we use more recent cites in the following--some of these cites are in this .tex file:} 
Classical examples including Q-learning, 
%\citep{watkins1992q},  \sam{would be nice for reader to provide a Q-learning cite that is easier to read and more recent} 
policy gradient algorithms, 
%\citep{sutton1999policy}, \sam{would be good to update prior policy gradient cite to a modern cite that uses regularization} 
and actor-critic methods \citep{sutton2018reinforcement}.
%\citep{konda1999actor} \sam{updated cites are often easier to read---}. 
%\YG{I find these standard methods proposed long ago, and for a reader to learn, the clearest reference would be a standard RL textbook.}
A recent influential model-free approach is Soft Q-learning \citep{haarnoja2017reinforcement, schulman2017equivalence} (and successors like Soft Actor–Critic \citep{haarnoja2018soft, haarnoja2018soft1, tan2025actor} and subsequent works \citep{hafner2019learning, ahmed2019understanding, geist2019theory, cai2020provably, zhan2023policy, ged2024matryoshka, lin2025optimistic}), which 
%\sam{would be good to use more recent cites as there is much work in this area currently.} 
incorporate entropy regularization for robustness and stability, and has seen broad empirical use \citep{raffin2021stable, hu2022soft, luo2025precise, anbazhagan2025adaptive}.
The second component, the \emph{decision-making} component, governs how actions are selected given the information learned so far and has also been extensively studied.
Simple yet fundamental strategies like $\epsilon$-greedy and Explore-Then-Commit naively separate exploration and exploitation: $\epsilon$-greedy algorithms mix random exploration with choosing the current estimated best action at each step, while explore-then-commit algorithms explore uniformly for a fixed period and then commit to the empirically determined best action \citep{langford2007epoch}.
Other widely used strategies that explore in arguably a more principled way follow an exploration-exploitation strategy based on the learning algorithm's uncertainty. Optimism in the face of uncertainty is one approach: for example,  a contextual bandit  agent maintains upper confidence bounds on the reward function and selects actions with the highest upper confidence bound \citep{ abbasi2011improved, filippi2010parametric, auer2008near, azar2017minimax}. This approach could, of course, be modified to be pessimistic (i.e., using lower confidence bounds). An alternative 
%that is neither optimistic nor pessimistic in the face of uncertainty 
is posterior sampling (e.g. Thompson Sampling): the agent selects actions according to their probability of being optimal under the posterior distribution \citep{strens2000bayesian, 
agrawal2012analysis, agrawal2013thompson, 
russo2016information, osband2019deep}; 
see \cite{russo2018tutorial} for a comprehensive tutorial.  

Despite this foundational work, a substantial gap remains between the theoretical performance of RL algorithms and their empirical performance \citep{laidlaw2023bridging}. One contributing factor is the limitation of regret as a performance measure. In practical deployments, the true regret is unobservable: it requires the knowledge of optimal value function, which is unknown. Consequently, it is difficult to benchmark algorithms by regret or compare them to regret bounds in practice. Moreover, many regret analyses focus on minimax bounds that maximize worst-case regret over a  class of environments. 
While useful for worst-case settings, these results are often overly conservative in many real-world  applications. 

In this section, we focus on another key challenge in real world RL---data scarcity (Challenge \textbf{C1}). In many applications in RL, limited interactions with the environment and the absence of high fidelity simulator lead to poor practical performance of RL algorithms; these issues are further exacerbated by missing data, user disengagement, and other operational constraints. Even when RL algorithms enjoy cumulative regret guarantees, they may still exhibit very poor performance in the early stages of learning, when data are limited. This is especially consequential in high-stakes settings such as healthcare, where early decisions can have strong impacts on individuals. For this reason, careful management of bias–variance tradeoffs is central to practical RL: introducing bias through structure, regularization, or information sharing can reduce variance and stabilize early decisions, whereas overly aggressive exploration may lead to highly variable and potentially nonsensical action selection. 
%Note that in real health applications the action space is usually constrained by the current state.  This is for ethical or feasibility reasons.  This is a problem with LLMs as it is hard to constrain the action space (aka via guardrails).
In addition, recent advances in large language models (LLMs) and foundation models offer new opportunities to improve online decision-making under data scarcity by leveraging pretrained representations and prior information from large external data sources. The remainder of this section discusses how the above ideas can be incorporated into online RL.

%, real RL deployments introduce additional complexities---such as data scarcity---that prior work does not fully address.

% \sam{is prior statement accurate?} \ziping{I changed to "many" because a significant body of literature focuses on Bayesian regret.} 
%ignore potentially very poor choices of actions for small $t$. \ziping{I don't get the part about ignoring poor actions for small $t$. } 
%\YG{my understanding is that many algorithm with theoretical guarantees only practically work well for $t$ very large}
%RL algorithm's reward and optimal policy's reward for a short period of time) and thus may not reflect typical real-world conditions. 
%Evaluation criteria that more closely reflect real-life performance of online RL are needed. 
\subsection{Bias-variance Tradeoff}
\label{sec:bias-variance}

% \sam{connect explicitly to challenge (1). } 
One major challenge in deploying RL algorithms in the real world is that many algorithms designed for gaming or simulator settings require very large datasets. However, we are interested in problems where the agent may not interact extensively with the target environment or a high-fidelity simulator to obtain a large amount of data, i.e., challenge \textbf{(1)} in Sec. \ref{sec_intro}. A typical example is when the primary goal is to interact with a human in a potentially critical setting. Even with a large dataset from a general population, an RL agent must quickly adapt to each new individual using only a few interactions, as discussed in Sec. \ref{sec_realWorld}. This motivates algorithms that perform well in a small-data regime. In statistics, strong finite-sample performance is often achieved via careful bias–variance tradeoffs---intentionally introducing bias (e.g. using simpler models, dropping less-informative features, or adding regularization) to reduce variance and improve performance 
%aka higher reward
with limited training data or interactions.

Common strategies for bias–variance control in RL include using learning algorithms that pool data across heterogeneous individuals, simplifying model structure and using a shorter learning horizon than the true horizon. The first strategy, pooling, refers to running an RL algorithm jointly across a group of users rather than separately for each individual, with decisions at each time point informed by the aggregated interaction history of all users (see Fig.~\ref{fig:pooling} for illustration). Pooling combines data from multiple individuals to reduce variance in early learning \citep{yom2017encouraging, figueroa2021adaptive,  piette2022artificial, nahum2024optimizing, coughlin2024mobile} but risks biasing policies toward the population average. Partial pooling methods, such as mixed-effects models \citep{tomkins2021intelligentpooling, ghosh2024rebandit}, aim to balance shared and individual-specific components. Besides pooling, simplifying model classes or restricting feature sets can also improve sample efficiency via bias-variance tradeoff \citep{ma2023sequential,westphal2024information}. Adjusting the discount factor is another common decision: using a smaller factor discounts future rewards, thereby reducing variance in cumulative returns but biasing the agent toward optimizing shorter-term outcomes \citep{sutton1998reinforcement, jiang2015dependence}. 
In the following, we first discuss the use of causal knowledge as a concrete and important mechanism for navigating bias–variance tradeoffs in RL. We then discuss additional considerations that generally arise in practical deployments for bias–variance tradeoffs, including issues related to model misspecification. 

\textbf{Causal RL.}
\begin{figure*}[t]
    \centering
    \begin{tikzpicture}[->, thick, main/.style={font=\sffamily}]
    \tikzstyle{fixedwidth} = [draw=none, text width=0.7cm, align=center]
    \matrix [column sep=0.5cm, row sep=0.25cm] {
        % & \node[fixedwidth] (O0) {$O_{d}$};
        % & & & & & & & & 
        % & \node[fixedwidth] (OK) {$O_{d+1}$}; \\
        \node[fixedwidth] (R00) {};
        &
        & \node[fixedwidth] (R0) {$R_{d-1}$};
        & & & & & & & & 
        & \node[fixedwidth] (RK) {$R_{d}$}; 
        % & & & & & & & & 
        % & \node[fixedwidth] (RRK) {$R_{d + 1}$}; 
        & 
        & \node[fixedwidth] (R6) {}; 
        \\
        & & &
        & \node[fixedwidth] (M1) {$M_{d, 1}$};
        & \node[fixedwidth] (M2) {\dots};
        & 
        & \node[fixedwidth] (MK) {$M_{d, K}$};
        % & & & & 
        % & & \node[fixedwidth] (MM1) {$M_{d + 1, 1}$};
        % & \node[fixedwidth] (MM2) {\dots};
        % & & \node[fixedwidth] (MMK) {$M_{d + 1, K}$};
        & \\
        \\
        & &
        & \node[fixedwidth] (C1) {$C_{d, 1}$}; 
        & \node[fixedwidth] (A1) {$A_{d, 1}$};
        & \node[fixedwidth] (A2) {\dots};
        & \node[fixedwidth] (CK) {$C_{d, K}$}; 
        & \node[fixedwidth] (AK) {$A_{d, K}$};
        % & & & & 
        % & \node[fixedwidth] (CC1) {$C_{d + 1, 1}$}; & \node[fixedwidth] (AA1) {$A_{d + 1, 1}$};
        % & \node[fixedwidth] (AA2) {\dots};
        % & \node[fixedwidth] (CCK) {$C_{d + 1, K}$}; & \node[fixedwidth] (AAK) {$A_{d + 1, K}$};
        & \\
        \\
        & & &
        & \node[fixedwidth] (N1) {$N_{d, 1}$};
        & \node[fixedwidth] (N2) {\dots};
        & 
        & \node[fixedwidth] (NK) {$N_{d, K}$};
        % & & & & 
        % & & \node[fixedwidth] (NN1) {$N_{d + 1, 1}$};
        % & \node[fixedwidth] (NN2) {\dots};
        % & & \node[fixedwidth] (NNK) {$N_{d + 1, K}$};
        & \\
        \node[fixedwidth] (E00) {};
        & 
        & \node[fixedwidth] (E0) {$E_{d-1}$};
        & & & & & & & & 
        & \node[fixedwidth] (EK) {$E_{d}$}; 
        % & & & & & & & & 
        % & \node[fixedwidth] (EEK) {$E_{d + 1}$}; 
        & 
        & \node[fixedwidth] (E6) {}; \\
    };
    \path[every node/.style={font=\sffamily}]
        % (R0) edge [color7] node [right] {} (O0)
        % (RK) edge [color7] node [right] {} (OK)
        (R00) edge [color1] node [right] {} (R0)
        (R0) edge [color1] node [right] {} (RK)
        % (RK) edge [color1] node [right] {} (RRK)
        (RK) edge [color1] node [right] {} (R6)
        (E00) edge [color6] node [right] {} (E0)
        (E0) edge [color6] node [right] {} (EK)
        % (EK) edge [color6] node [right] {} (EEK)
        (EK) edge [color6] node [right] {} (E6)
        (E0) edge [color6, bend right=45] node [right] {} (RK)
        (R0) edge [color1, bend left=45] node [right] {} (EK)
        % (EEK) edge [color6] node [right] {} (RRK)
        (R0) edge [color2] node [right] {} (M1)
        (R0) edge [color2] node [right] {} (MK)
        (M1) edge [color3] node [right] {} (RK)
        (MK) edge [color3] node [right] {} (RK)
        (E0) edge [color4, bend right=15] node [right] {} (M1)
        (E0) edge [color4] node [right] {} (MK)
        % (R0) edge [color2] node [right] {} (N1)
        % (R0) edge [color2] node [right] {} (NK)
        % (N1) edge [color3] node [right] {} (RK)
        % (NK) edge [color3] node [right] {} (RK)
        (E0) edge [color4] node [right] {} (N1)
        (E0) edge [color4] node [right] {} (NK)
        (N1) edge [color5] node [right] {} (EK)
        (NK) edge [color5] node [right] {} (EK)
        (A1) edge node [right] {} (M1)
        (AK) edge node [right] {} (MK)
        (C1) edge node [right] {} (M1)
        (CK) edge node [right] {} (MK)
        (A1) edge node [right] {} (N1)
        (AK) edge node [right] {} (NK)
        % (C1) edge node [right] {} (N1)
        % (CK) edge node [right] {} (NK)
        % (RK) edge [color2] node [right] {} (MM1)
        % (RK) edge [color2] node [right] {} (MMK)
        % (MM1) edge [color3] node [right] {} (RRK)
        % (MMK) edge [color3] node [right] {} (RRK)
        % (EK) edge [color4, bend right=15] node [right] {} (MM1)
        % (EK) edge [color4] node [right] {} (MMK)
        % (R0) edge [color2] node [right] {} (N1)
        % (R0) edge [color2] node [right] {} (NK)
        % (N1) edge [color3] node [right] {} (RK)
        % (NK) edge [color3] node [right] {} (RK)
        % (EK) edge [color4] node [right] {} (NN1)
        % (EK) edge [color4] node [right] {} (NNK)
        % (NN1) edge [color5] node [right] {} (EEK)
        % (NNK) edge [color5] node [right] {} (EEK)
        % (AA1) edge node [right] {} (MM1)
        % (AAK) edge node [right] {} (MMK)
        % (CC1) edge node [right] {} (MM1)
        % (CCK) edge node [right] {} (MMK)
        % (AA1) edge node [right] {} (NN1)
        % (AAK) edge node [right] {} (NNK)
        % (C1) edge node [right] {} (N1)
        % (CK) edge node [right] {} (NK)
        ;
    \end{tikzpicture}
    \caption{An example of an early causal DAG used to describe the  environment in preparation for the fifth deployment of the HeartSteps JITAI, (see Section \ref{sec_realWorld}). The action $A_{d, k}$ refers to whether a walking suggestion is sent or not. Each day $d$ contains $K$ interventions. The reward $R_d$ is the daily commitment to being active. The engagement $E_d$ represents the user's daily engagement with the app. The context can include the current activity status of the user, the location of the user, etc. The variables $M_{d, k}$ and $N_{d, k}$ are the mediators between $A_{d, k}$ and $R_d, E_d$.}
    % \sam{either make the figure general or make it specific to heartsteps.  right now you have mixed the two with K arbitrary and calling a bag by the term "day."  I suggest you make this DAG general so then it is a "Causal DAG for bag $d$"}
    \label{fig:dag}
\end{figure*}
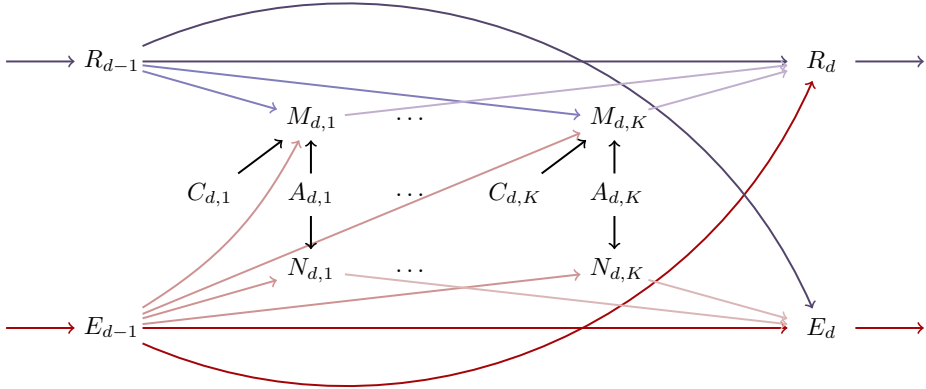

% \daiqi{Start with descriptions about causal RL and sample efficiency. Organize w.r.t. open questions: misspecified causal DAG, real-life deployment. Move the literature review under each open question, but focus on the assumptions in the DAG and deployment. Can cite the same paper multiple times, but should focus on different aspect.}
The use of causal knowledge provides an additional approach to trading bias (potentially incorrect causal knowledge) with variance (reduction in complexity provided by the causal knowledge). This tradeoff is directly relevant to challenge \textbf{C1}: when each individual or deployment yields only limited interactions with the environment, a causal model can restrict what the RL algorithm needs to learn from scratch.
% \kwz{I feel the following sentences do not make it crystal clear how causal reasoning actually helps improve sample efficiency.} 
%\textcolor{red}{For example, domain scientists can depict causal relations among variables using a causal DAG.}
%\textcolor{red}{Causal RL can use DAGs that go beyond the standard RL DAGs in Fig.~\ref{fig:rl_dag} by encoding problem-specific knowledge, such as which parts of the history are relevant for future rewards and which proximal variables mediate effects on distal outcomes.}
%\textcolor{red}{Fig.~\ref{fig:dag} is an example causal DAG constructed for the next deployment of the HeartSteps JITAI (see  Secs.~\ref{sec_realWorld} and~\ref{sec:HeartSteps}) \citep{gao2025harnessing}.} 
% Simpler examples of causal DAGs are the DAGs shown in Fig. \ref{fig:rl_dag}. 
% \sam{Daiqi, would you eliminate the arrow from E to R in this DAG?} 
%This DAG is a simple version of the DAG that was collaboratively developed with  domain scientists.
% \YG{Unclear whether the intended meaning here is “represent” or “provide.” Please clarify the wording of this sentence.}. 
%The key information provided by a 
Causal DAGs such as the one shown in Fig. \ref{fig:dag} encode conditional causal independencies among sets of variables. The \textit{d-separation rules} provide a systematic method to visually interpret these conditional independencies from the causal DAG \citep{pearl2009causality}. 
%These conditional independencies can be utilized as inductive biases when designing an RL algorithm to  help engineer reward signals or help reduce the number of parameters that are learned by identifying necessary state variables.
% \asim{Let's add a reference on d-separation}
%\sam{To my knowledge "causal RL" is really about off-policy RL using a batch of observational data.  -----Bareinboim's work is definitely under that topic.  Need to check the others. --- Also there is  somebody who has done a lot in this area--I don't think it is Bareinboim so  we need to find this person...---Our paper is about online RL in which the RL agent can explore so even if we use causal DAGs we are not doing causal RL!}  
For comprehensive reviews of causal RL,  see  \cite{deng2023causal,zeng2024survey,correa2024ctfcalc,ge2025review}.
% \kwz{I feel the flow between the following paragraphs could be improved. Perhaps make an outline here of what the reader should expect will be covered in this section? I feel like its a long list of things right now, and could be more logically organized.} \kwz{I think this section could also potentially be made shorter to only contain most important points; see some comments below.}
%\YG{Can we shorten this subsection further? One option is to group papers that address similar topics and avoid going into excessive detail. We can focus on highlighting the key points and making them clear.}
  
Causal DAGs can address data scarcity in three concrete ways: (1) constructing more informative rewards when the original reward is distal or sparse, (2) identifying state variables and action components most relevant to rewards and future states, and (3) embedding prior causal knowledge into Bayesian priors, counterfactual data, or invariant dynamics. We briefly summarize the work in each of these three areas.
%to highlight the need to account for an incorrectly specified causal DAG. 

\textit{Distal, Sparse Rewards.}
Online RL is often challenged by the lack of feedback following each decision, which can occur when the scientific goal is to optimize a long-term or distal outcome (e.g., an individual's overall cardiovascular health after six months) or sparse outcome (e.g. intermittently collected survey data from an individual).
%In these settings there may be a  meditational pathway from action to sparse and/or distal outcomes (e.g., HeartSteps's intervention influences a user’s affective association with physical activity through a long mediational pathway involving short-term bouts of physical activity).  Even when there is a domain science specified meditational pathway,  readily observed short-term outcomes or mediators may  be \textit{noisy} and thus . 
To address these challenges, a variety of algorithms have been proposed to design improved rewards based on proximal outcomes that are causally related to the distal outcome.  These methods build off of classical statistical methods including  \citep{prentice1989surrogate,weintraub2015perils}. 
One line of work uses surrogate methods \citep{athey2019surrogate} to construct a function of (multiple) proximal outcomes that serve as a substitute for the distal outcome \citep{yang2024targeting,tripuraneni2024choosing,wang2022surrogate,zoucausal}.
% One line of work uses surrogate methods to construct a function of (multiple) proximal outcomes that serve as a substitute for the distal outcome. \cite{yang2024targeting} utilize historical data to construct a surrogate index \citep{athey2019surrogate} and deploys the surrogate index in a contextual bandit environment using an online Thompson sampling contextual algorithm \citep{russo2018tutorial} to optimize for distal user retention in a study with the Boston Globe. Similarly, \citep{tripuraneni2024choosing,wang2022surrogate} also use existing data to construct and deploy a surrogate outcome for use in  A/B experiments and in online RL, respectively.
%Other works leverage historical data, potentially from previous A/B testing experiments, to learn what short-term outcomes might be an effective proxy metric for long-term outcomes \citep{tripuraneni2024choosing,wang2022surrogate}. 
%The above methods constructed the reward using domain knowledge or offline data, fixing the definition during deployment. 
% \citet{zoucausal} provide regret bounds illustrating that, when the surrogate reward is learned online, a contextual bandit can obtain lower regret by using the surrogate reward as compared to using the original reward.
Other approaches for optimizing for long-term outcomes in the literature do not {replace} the long-term outcome with a short-term surrogate but rather combine proximal outcomes with distal outcomes \citep{wu2022partial, zhang2025impatient, anderer2022adaptive}. 

%For example, \cite{wu2022partial} develop a Thompson sampling algorithm that can accommodate incremental, censored survival outcomes, and \cite{zhang2025impatient} develop a Thompson sampling based algorithm that incorporates progressive feedback revealed over time (e.g., user engagement across multiple days). Additionally, \cite{anderer2022adaptive} posit a joint Bayesian model over intermediate and distal outcomes, and use the intermediate outcome as a measure of an intervention's effectiveness to decide whether to stop a trial early.

% \textit{Large State and/or Action Spaces.}  
% ********** \sam{this section is highly repetitive. we need to combine and shorten greatly...}
\textit{Defining States and Selecting Actions.} 
Some algorithm design focuses on using causal reasoning to simplify learning for \textit{large state} and/or \textit{large action spaces} (which pose challenges for learning due to the curse of dimensionality) by focusing on the \textit{parents} of the reward and next state in a DAG. In a DAG, an arrow from A to B implies that A is a parent of B. 
% \kwz{Don't we already have a section on POMDPs and unknown state? Maybe should merge the following with that section?} \daiqi{We focus on the definition of states using causal reasoning regardless of whether the states are observed or not, instead of learning the (pre-defined) latent states.}
% Even when the state space in a naturally defined problem is large, a subset of the state might be enough to satisfy the modeling assumptions (e.g. Markov transitions) and learn the optimal policy. 
% A causal DAG indicates which subset of the states or actions is independent of future rewards given the rest, and thus can be omitted. 
In a stationary MDP, a minimal state or action would trivially consist of the parents of the reward and the next state. 
% \sam{Daiqi I used adjective, "minimal."  Is this inaccurate?} \daiqi{Yes it's accurate.}
%Ian Osband and Benjamin Van Roy. Near-optimal reinforcement learning in factored mdps. In Advances in Neural Information Processing Systems , 2014
%Malcom Strens. A Bayesian framework for reinforcement learning. In Proceedings of the 17thInternational Conference on Machine Learning , pages 943–950, 2000.
Nevertheless, causal DAGs are particularly useful for state construction in a history-dependent environment (see Sec.~\ref{sec_notation}), see \citet{zhang2020designing} for finite-horizon problems and \citet{gao2025harnessing} for bagged decision times.
In addition, model based RL algorithms may take advantage of a DAG-based factorization of the state transition probabilities, this leads to the large area of \lq\lq Factored"  MDPs, see \citet{osband2014near} for an example.
In action selection, even when direct intervention on these parents is infeasible, knowledge of intervention effects can be shared across interventions since all effects go through the same parents \citep{lu2020regret,lu2022efficient}. 
Causal bandits is another line of work that aim to learn the optimal policy for bandits with many actions and under varying assumptions \citep{bilodeau2022adaptively,feng2023combinatorial,varici2023causal,yan2024causal}.
% ************************* \sam{reorganize condense above; please comment on whether the method was implemented online in real life....}
% In both \cite{lu2020regret} and \cite{lu2022efficient}, the benefit comes from the fact that the size of the parents of the reward (and the next state) is smaller than the size of the full action space.
% \kwz{Why do we care about designing loss functions? How do the methods described improve exploration? I feel it is all too vague. like what is "causal impact"?} 
% Another line of work aims to leverage causal knowledge to improve exploration by prioritizing certain states or actions.
% , or to increase exploitation efficiency by reusing observed data or derived quantities in the training process.
\citet{seitzer2021causal} propose to improve exploration by prioritizing certain states or actions. 
% \citet{seitzer2021causal} proposed three approaches to improve exploration efficiency based on the causal action influence (CAI), defined as the conditional mutual information between an entity's next state and the action given the current state. 
% They proposed three approaches to increase sample efficiency: adding CAI as a reward bonus for visiting states of high CAI, encouraging exploration of actions with higher CAI, and prioritizing replay of episodes with higher CAI during training.
% \sam{please comment on whether any of the  methods in above paragraph has been implemented online in real life!}
% \asim{The above involves quite a bit of densely written jargon. Can we write this more accessibly? Or can we summarize their key insight in using causal and counterfactual reasoning and skip some details? Also, our preliminaries do not even use the term Q value so maybe we should just use longer explanations or terms like action value functions and such just to keep the writing relatively self contained.}

\textit{Bayesian Priors, Counterfactual Data Augmentation, and Invariance Dynamics.}
%Causal information may be incorporated as prior knowledge through Bayesian \textit{priors}, \textit{counterfactual data augmentation}, or \textit{invariant dynamics}. 
% \sam{1-2 sentences discussing anna/susobhan papers about putting informative priors on parameters in the reward function in contextual bandit.  Explain how these priors based on existing data and domain expertise and provide cites.}
Leveraging domain knowledge about the causal relationships between states and rewards, priors for the policy parameters can be constructed by fitting models to existing data \citep{liao2020personalized, trella2023reward, ghosh2024rebandit}.
% we already wrote this: This approach aims to increase sample efficiency by leveraging prior knowledge about the causal structure of the environment to speed up learning.
%Rather than placing the prior directly on policy parameters as in standard posterior sampling, 
More generally, \citet{mutti2024exploiting} use a partially specified causal DAG to define hierarchical priors over possible DAGs and their factored transitions. 
% The posterior distribution for the underlying causal DAG is then updated online throughout learning as the belief about the most likely DAG of the possible DAGs becomes more certain. 
% \sam{tell reader whether this method has been implemented in real life.}
% Offline, but assumes the the standard MDP causal DAG
% \citet{lu2020sample} modeled the state transition dynamics as a SCM, which is estimated from data using a generative adversarial network (GAN)-like adversarial framework, then generate counterfactual data given alternative actions to augment the dataset.
% Offline, need to learn the causal structure
% \citet{sun2024acamda}
Counterfactual data augmentation involves estimating parameters in models based on a known causal DAG for use in generating counterfactual data. This counterfactual data can then be used  to artificially increase the total sample size in online learning \citep{buesing2018woulda,chen2023intrinsically}.  
In fact, counterfactual data augmentation is, in special cases, equivalent to use of a prior in a Bayesian algorithm, as demonstrated in bootstrapped Thompson sampling \citep{osband2015bootstrapped}.
%Thompson sampling, or posterior sampling, is a principled exploration method in which the agent samples the optimal action from its posterior distribution, combining prior knowledge with observed data.
%Counterfactual data augmentation can also be used to approximate the  posterior distribution  in Bayesian RL.
%can be intractable, creating computational challenges when implementing the algorithm.
Bootstrapped Thompson sampling combines artificial data generated from the prior distribution with the observed data and then bootstrap samples from the combined data. 
% These bootstrap samples are used by the online learning algorithm.
Finally, causal knowledge enables generalization across different environments or tasks by learning invariant causal dynamics, adapting only the varying parts \citep{zhang2017transfer,sontakke2021causal}.

%, so that the posterior distribution is equivalent to the combination of artificial and observed data.

% \citet{mendez2023carl} learn the causal model using score-based causal discovery algorithms, before filtering actions that likely lead to negative rewards and prioritizing actions likely lead to positive reward.
% \textbf{Open Questions.}
In the above approaches, only the parameters or strengths of the causal links need to be learned, either explicitly via model-based methods or implicitly via model-free methods. 
However, a perfectly specified causal DAG is a strong assumption, especially across changing deployments.
An important open question is how to construct algorithms that leverage inductive biases from the causal DAG to efficiently learn, yet optimally trade off bias and variance. %are robust to some degree of mis-specification of the causal DAG. 
A possible approach is to separate the causal assumption-based model (e.g. factored transitions) from the model-free RL algorithm \citep{gao2025active}.
See also Sec.~\ref{sec_nonstationary_continual} for a discussion of revising causal DAGs through continual learning when challenge (2) arises.
% \sam{should prior sentence be connected to challenge (2) from intro and Figure 1? "(2) the presence of significant changes in the environment that require re-optimizing the decision making policy by the RL algorithm during next deployment and/or updating the RL algorithm for use the next deployment."}
% There remain open questions in developing methods to construct surrogate outcomes for rewards that have weaker assumptions, and developing online decision-making algorithms that are able to practically perform well when learning from distal noisy rewards.

% \textit{Real-World Deployment of Causal RL.} 
%\daiqi{This paragraph currently focuses on causal RL. Not sure if it will be a common challenge for other topics as well.}
Causal RL is promising for data-scarce applications, but it is not yet routine in real-world deployments.
To our knowledge, there has not yet been a real-world deployment of an online RL algorithm that explicitly leverages causal DAGs to improve learning efficiency.
%\textcolor{red}{Existing human-facing online RL deployments have often used simpler, prespecified models to ensure stability and meet clinical or operational constraints \citep{yom2017encouraging, trella2023reward, ghosh2024rebandit}.} 
This history does not diminish the motivation for causal RL; rather, it highlights what is needed for causal DAGs to be used responsibly in deployment: clearly stated and justified DAG assumptions, sensitivity analyses, and algorithms that translate domain knowledge into measurable gains in sample efficiency. 
Developing and evaluating such deployment-ready causal RL methods is therefore a crucial open problem.

\textbf{Misspecification.}  %\sam{explicitly connect to challenge (1) in intro.}
Model misspecification is common in sequential decision-making, particularly in complex environments where data is limited and noise is substantial (Challenge \textbf{(1)}). One reason is that in many applications, the agent typically does not know which model class the state dynamics and/or true reward process belongs to \citep{dulac2021challenges}. A growing literature, largely theoretical, examines decision making when the state, reward, or value function models are misspecified, typically asking how model error affects the achievable reward or regret \citep{ghosh2017misspecified, lattimore2020learning, foster2020adapting, zanette2020learning, krishnamurthy2021tractable, krishnamurthy2021adapting, yang2021bandit, guo2024online}.

Here, we emphasize a less-noticed but practically important source of misspecification that is closely tied to bias–variance tradeoffs: in many applications, RL designers often intentionally adopt a simpler---but knowingly misspecified---reward model for more stable exploration \citep{yom2017encouraging, trella2022designing, ghosh2024rebandit, zhang2025replicable}. While more complex models may reduce bias, they can also amplify variance. Such effects are more consequential in RL than in standard non-adaptive supervised learning, because misspecification can affect policy stability, autonomy, and the replicability of downstream or after-study analyses \citep{zhang2025replicable, guo2025statistical}. Here, policy instability means that, conditional on the state, the action-selection probabilities fail to converge as data accrue---e.g., they oscillate or converge to different subsequence limits as the number of trajectories or the time horizon grows. For example, \citet{guo2025statistical} show that when a linear contextual bandit algorithm (LinUCB; \citealp{abbasi2011improved}) is applied in an environment with nonlinear reward structure, the action selection probabilities conditional on the same state may fail to concentrate as time progresses and instead diverge across repeated runs of the same algorithm on the same environment. One downstream implication is that standard statistics, such as the ordinary least squares estimator, may not be replicable, as they may fail to converge across runs or yield interpretable results. As another example, \citet{zhang2025replicable} study a longitudinal setting in which a single RL algorithm pools data online across a  group of  users. They show that when standard RL algorithms that pool data across users to learn nonsmooth policies, such as $\epsilon$-greedy and Thompson sampling, are combined with misspecified models, the probabilistic policy for an individual user may not converge to a deterministic limit even as the number of users tends to infinity, which also leads to failures in downstream inference. These examples underscore the importance of designing online RL algorithms that remain stable and reliable under model misspecification, and of developing principled strategies for managing bias–variance tradeoffs in practical deployments.

\textbf{Toward a unified framework for bias–variance tradeoffs in RL.} Finally, we point out that one key challenge lies in the absence of a unified framework for understanding and characterizing the bias-variance tradeoff. Many bias–variance tradeoff mechanisms are theoretically connected. For instance, \cite{rathnam2024rethinking} showed that discount factor regularization is equivalent to imposing a uniform prior on the state transition model, effectively regularizing its complexity. This equivalence suggests that one regularization decision may have unintended effects on others, making it difficult to navigate among various decisions in practice. \cite{arumugam2022rate} proposed a promising step toward a unified framework by defining a simpler learning target that adapts complexity based on observed sub-optimality gaps, though its current Bayesian formulation currently limits practical implementation.

%???\YG{I don't get how this paragraph is connected to text up and below} There is also a need for systematic guidance on hyperparameter tuning and model selection in data-scarce domains such as healthcare (see Sec. \ref{sec_realWorld}). While recent work in deep RL \citep{patterson2024empirical} offers empirical advice on design choices and algorithmic stability, this guidance does not readily extend to data-limited, heterogeneous environments like those encountered in real-world interventions on humans.

When faced with multiple strategies that embody different bias–variance tradeoffs, the RL designer is often unaware of the optimal choice prior to deployment. Beyond using offline data or simulators for comparison (Sec. \ref{sec_sequential_design}), this challenge motivates research on online model selection \citep{dann2024data}, which adaptively identifies the best-performing design during ongoing experimentation. Formally, this involves a collection of algorithms mapping histories to policies, with the goal of achieving performance competitive with the best algorithm in hindsight. Recent theoretical approaches estimate regret scaling coefficients that capture each learner’s behavior \citep{dann2024data,krishnamurthy2024towards}, attaining regret bounds comparable to the best base learner. 
Other theoretical work \citep{cutkosky2021dynamic, pacchiano2020regret, pacchiano2020model}  relaxes assumptions to improve adaptability and robustness. How these works might guide practical real-life use, particularly in human environments, is not yet clear.
%However, these approaches build upon a given candidate algorithm set---with few discussions on how to construct a suitable set---potentially limiting their practical applicability. 

% Despite these strategies, selecting an appropriate bias–variance balance remains largely empirical. Current practices rely heavily on ad hoc tuning and heuristics that are difficult to justify statistically or replicate across studies. Several open challenges remain:

%\ziping{left out POMDP as there are too much cream in this paper}
% \input{POMDP}

\subsection{Online learning with Large Language Models (LLMs)} 
\label{sec_LLM}
The challenges discussed above fundamentally stem from limited data, which forces compromises such as overly simple or misspecified models and restrictive bias–variance tradeoffs. In many applications involving humans, rich semantic background information, such as pretreatment information in electronic health records, and immediate, high-context data, such as student question-answer interactions in online education, or browsing behavior in digital marketing—are often available. However, this information is typically unstructured (e.g., text) and high-dimensional, making it difficult to incorporate into traditional online RL algorithms. As a result, the agent often learns from scratch from only coarse summaries of this information. Large language models (LLMs), which encode broad knowledge from natural language, offer a way to bring this background and textual information directly into the learning process.

By analyzing rich contextual information, such as a user's profile or brief interactions, an LLM can infer unobserved states, effectively mitigating data scarcity. 
%partial observability and model misspecification described above. 
For instance, in health intervention settings, an LLM can read a participant's open-ended self-report (e.g., “my leg is sore today”) and use it to adjust or veto the RL agent's recommended action---information a text-blind policy would otherwise have to recover as a latent state from limited sensor monitoring under a partially observable MDP \citep{karine2025using}. 
% \sam{ need to clarify what  "a standard RL policy would otherwise treat as latent" means... I think we are talking about a POMDP with improved observations of latent state?} 
Such approaches have shown some promise across domains, including inferring latent depression \citep{ohse2024zero}, loneliness \citep{garg2023lonxplain}, and pain \citep{amidei2025exploring} from text that would otherwise have been difficult to infer from the non-textual data collected. In the social and behavioral sciences, similar techniques have been used to trace students' latent knowledge states in online education \citep{zhan2024knowledge}, segment customers in marketing \citep{li2025consumer}, and infer relevance in e-commerce search \citep{soviero2024chatgpt}. Some have been validated with human participants, such as \cite{li2023eliciting} on content preferences, moral reasoning, and information validation, and \cite{cosentino2024towards} on subjective sleep health.
% \sam{has any of this been used in real-life? if so we should explicitly state this.}

% \textbf{Combine LLM with RL.} \cite{karine2025using} combine LLM with RL agent (Thompson Sampling): RL first select a candidate action at each time step. Then based on the LLM prompt that includes the user preference and other information, the LLM decides whether to deploy the RL candidate action. Some other work \citep{verma2025balancing} use LLM as a reward design agent that provides reward signals based on rich contextual information that has been shown to outperform static, manually crafted reward signal. 
A more ambitious approach avoids using a separately trained RL agent and instead treats the LLM itself as an in-context sequential decision-making agent: the pre-trained model is prompted with the current state and interaction history, then directly outputs the next action \citep{yao2022react, monea2024llms}. Because the model's parameters are never updated \citep{brown2020language}, the interaction history acts as context that guides successive decisions rather than as training data. \cite{yao2022react} make this concrete by prompting the model with explicit \textit{Reason} and \textit{Act} steps, and show that such a GPT-based agent can outperform traditional RL agents on interactive simulated benchmarks such as ALFWorld \citep{shridhar2020alfworld} and WebShop \citep{yao2022webshop} with only a few demonstrations and no additional training.
% \sam{what do we mean by "recent studies"--do we mean in theoretical papers or in simulations?  Or do we mean in real-life studies.  Need more precision here!} 
   
% \sam{tell reader if this is in real life application or in a simulation environment.}

Applying the \textit{Reason}--\textit{Act} paradigm \citep{yao2022react} to sequential decision-making with real humans is only beginning to emerge. The closest instances are clinical diagnostic agents that gather information over multi-turn dialogue \citep{tu2025conversational} and iteratively order tests to refine hypotheses \citep{nori2025sequential, baniharouni2025language}, but these are evaluated on simulated patients and benchmark cases rather than deployed. No study has yet had an LLM explicitly reason and act over a predefined action space in a live online loop; the closest deployments are open-ended chatbots used with real users---tutors \citep{wang2024tutorcopilot, kestin2025aitutoring}, a therapy chatbot \citep{heinz2025randomized}, a health coach \citep{jorke2025bloom}, a recommender \citep{sun2024movie}, and a political-discussion assistant \citep{argyle2023leveraging}---alongside a parallel line evaluated offline or in simulation \citep{scarlatos2025training, chiu2024computational, ong2024advancing, feng2025expectation}. Such systems personalize effectively, but they hold open-ended conversations that demand heavy safety guardrails and do not exploit LLM reasoning for long-term outcome optimization. Real deployments and practical guidelines are therefore still needed before an LLM agent can replace the use of RL with in-context learning that optimizes long-term outcomes without giving incorrect or harmful feedback.

\section{Between deployment  learning and statistical inference}
\label{sec_post_trial}
% \kwz{(1) the inability to interact extensively with the target environment or a high-fidelity simulator to obtain a large amount of data and (2) the presence of significant changes in the environment that require re-optimizing the decision making policy by the RL algorithm during next deployment and/or updating the RL algorithm for use the next deployment.}
% \sam{ensure that this section makes explicit connections  to the two challenges in intro.} 
As shown in Fig. \ref{fig_intro}, another use-inspired avenue of research in RL involves statistical inference based on adaptively collected data from a prior (or ongoing) deployment, that is the actions were determined by a %(possibly unknown) 
\textit{behavior policy} (the policy induced by an online RL algorithm used to collect the data). 
%The term behavior policy is used to describe the policy that was used to interact with the environment in an online setting to collect data (i.e., store away the history). 
We split the objectives of what we call between-deployment learning and inference into two categories of offline learning and analysis: (i) policy learning, e.g., to learn an optimal policy from data, and (ii) statistical inference, e.g., assess causal treatment effects or the value of a policy. Both of these objectives closely support the redesign of an RL algorithm using previously collected data (e.g., from prior deployment) in preparation for subsequent deployment.

% \sam{I think  the cites in this  paragraph concern data from one long trajectory that follows an MDP. Please check.  We need to make the setting  clear to reader.  } 
\textbf{Policy learning.} Policy learning aims to estimate a near-optimal policy using data collected during the deployment of an RL algorithm. Much of this literature has focused on MDP settings, e.g., see \citet{levine2020offline} for a review. A central principle is conservatism under limited support: when the behavior policy rarely (or never) visits certain state–action pairs, guard against over-extrapolation by either optimizing a lower-confidence (pessimistic) value estimate or regularizing the learned policy toward the behavior policy to mitigate distribution shift \citep{fujimoto2019off, kumar2020conservative, kidambi2020morel}. %A complementary thread is safe/constrained policy optimization, which imposes explicit safety, cost, or baseline-improvement constraints so the learned policy is provably no worse than a reference policy \citep{achiam2017constrained}. 

% \sam{I think  the cites in this  paragraph concern data from one long trajectory that follows an MDP. Please check.  We need to make the setting very clear to reader! I think these cites permit the behavior policy to be an RL policy? } 
\textbf{Statistical inference.} Much research concerns the setting in which  i.i.d. trajectories have been collected by a known behavior policy in an MDP or longitudinal, non-Markovian environment. This type of data occurs when the RL algorithm learns and selects actions separately on each unit (no online pooling of data as discussed in  Sec.~\ref{sec:bias-variance} and~\ref{sec:inference-pooling}).  
This is the adaptive intervention setting.   There has been some limited work on statistical inference for parameters in statistical models in this area; see  \citep{boruvka2018assessing} for an example.   But the most common inferential goal 
 is off-policy evaluation (OPE), that is to estimate the value of a target policy using data collected under a (possibly different) behavior policy. 
Common OPE approaches  include (i) \emph{importance sampling (IS)}, which reweights rewards by the ratio of the target policy’s action probability to the behavior policy’s action probability so that sum of weighted rewards approximates the target-policy expectation \citep{uehara2022review,precup2000eligibility}; %li2011unbiased
(ii) \emph{model-based/direct methods}, which learn a value/transition or Q-function model from the data and then plugs in the target policy to compute its value \citep{sutton2018reinforcement}; and (iii) \emph{doubly robust} estimators, which combine the two approaches, using a model-based estimate and  an IS correction; these methods produce consistent estimators of the sum of rewards if either component is correct \citep{ jiang2016doubly,dudik2011doubly}. See \citet{uehara2022review} for a review of this area. Most results in this area assume the behavior policy is known.  If the data is observational, thus the behavior policy is unknown, handling unmeasured confounding is crucial, with recent advances using mediators, instruments, and bridge functions in confounded MDPs/POMDPs \citep{shi2022minimax,xu2023instrumental,yu2024two,shi2024off}.

% \sam{is the following about data from one long trajectory, assumed to follow an MDP OR is it about longitudinal data.  Actually I think we have some cites from each setting.  This needs to be made very clear to reader. These areas need to be separated as it is too confusing to mix.} \kwz{Daiqi, I'm not sure if this paragraph is about policy learning or statistical inference. Or both, so I'm not sure exactly how to address Susan's comments.} Another statistical line constructs estimators of policy performance---such as long-run cumulative reward or average reward---using doubly robust or efficient-influence-function scores with learned nuisance components (e.g., Q-functions, proximal bridge functions), and then optimizes over a policy class to maximize this estimated performance. This often yields identified and efficient procedures in longitudinal/health settings \citep{nie2021learning, liao2022batch, shi2024statistically, zhou2024estimating}.

%\textbf{Statistical inference: After adaptive experimentation.} 
There is a growing body of work on statistical inference using data collected from adaptive experiments; adaptive experimentation results in  non i.i.d. data. Parameters of interest include marginal treatment effects, causal excursion effects, and parameters in outcome models. Much of this literature focuses on adaptive experimentation in the  decision making horizon $H=1$ setting, specifically for  
bandit environments, including batched bandits, multi-armed bandits, and contextual bandits \citep{deshpande2018accurate, zhang2020inference, hadad2021confidence, zhang2021statistical, zhang2022statistical,  zhang2025replicable, bibaut2021post, guo2025statistical, leiner2025adaptive} (the optimal policy for $H=1$ settings is discussed in Section~\ref{common_environments}).  
Adaptive experimentation is also used in settings with horizon $H>1$; these are settings in which the RL algorithm interacts with a sequence of horizon $H$ episodes and  usually RL learning  and updating of the policy only occurs between the $H$ length episodes.  
\citep{ syrgkanis2023post} consider statistical inference  in this more general setting.
In all of the above adaptive experimentation  settings (regardless of whether $H=1$ or $H> 1$), if the policies used to select actions in each episode were prespecified, the data would be independent across the episodes  as each $H$ length episode concerns a different unit sampled from the population. However, since the bandit algorithm uses data from prior episodes to assign actions, the data collected by these algorithms are no longer independent, thus new methods are needed.

While the methods above provide many essential tools for statistical inference and offline policy learning, notable gaps remain within our framework. The subsections that follow highlight emerging directions we find important for advancing both methodology and practice.

%While the methods described thus far in this section provide essential tools for statistical inference and offline policy learning, several important challenges remain. The following subsections highlight two emerging directions—inference with longitudinal data collected using pooling algorithms (Fig. \ref{fig:pooling} and offline algorithm selection—that we view as especially important for advancing methodology and practice.

\begin{figure}[t]
    \centering
    \includegraphics[width=0.7\linewidth]{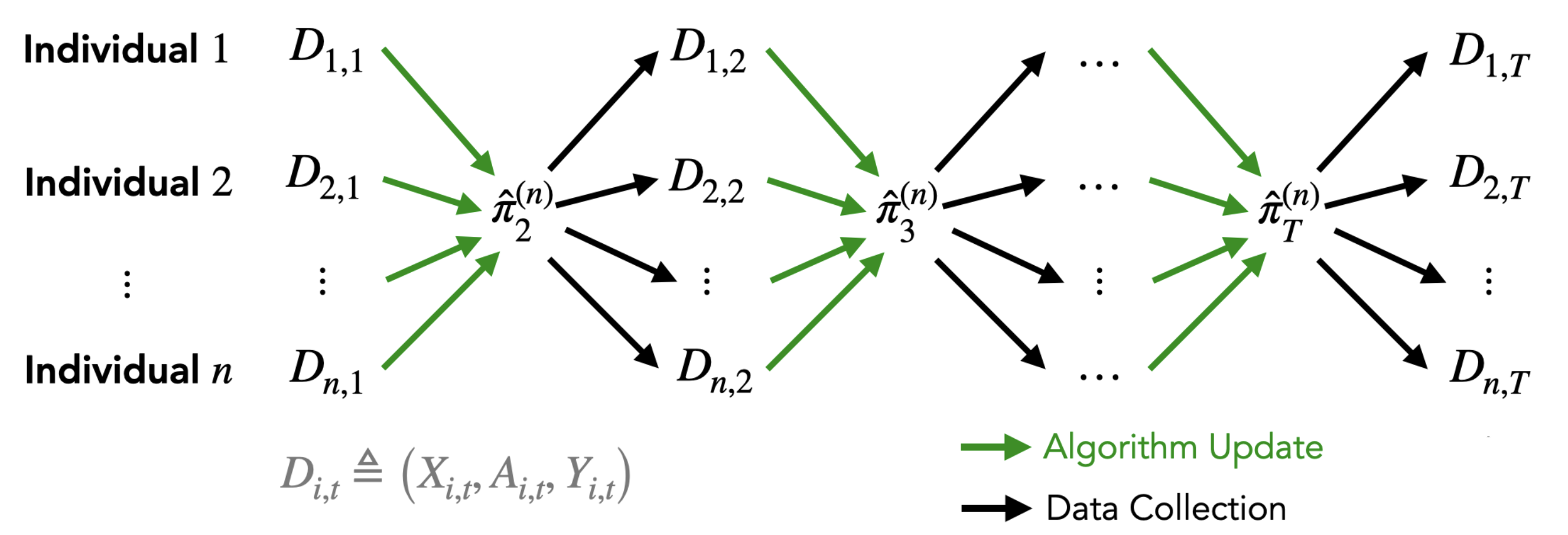}
    \vspace{-4mm}
    \caption{\textbf{Pooling algorithm.} %\sam{lets omit the name of the algorithm--this pooling problem has nothing to do with whether th RL alg is a bandit or a full RL alg.} 
    Depiction of how a pooling  algorithm combines or ``pools" data across individuals to form and update the policies over time. Adapted from \citet{zhang2024replicable}. \vspace{-3mm}}
    \label{fig:pooling}
\end{figure}

\begin{comment}
\begin{itemize}
    % \item overview - using previously collected data to learn things (e.g. treatment effects, estimate value, learn a policy, ect). 
    \item Estimate a quantity of interest (treatment effect, or value) - Kelly  (Talk about philosophy of not wanting to assume environment is "correct" to be valid - Kelly)
    \begin{itemize}
        % \item Off-policy evaluation
        % \item Inference on adaptively collected data besides OPE
        % \item excursion effects
        \item Observational data vs known behavior policy (Daiqi)
    \end{itemize}
    \item To learn a policy - Daiqi
    \begin{itemize}
        \item Policy optimization (using pessimism), batch policy learning
        \item Dynamic treatment regime
        \item Observational data vs known behavior policy
    \end{itemize}
\end{itemize}
\end{comment}

% \subsection{Inference based on  Data Collected from the Deployment of an Adaptive Intervention in which the RL Algorithm Pooled Data across Individuals}
\subsection{{Inference based on data collected from an online RL deployment}}
\label{sec:inference-pooling}
% \sam{note that this is NOT data from an adaptive experiment --we have to ensure that we don't let reader get confused!}
%In several applications such as the HeartSteps JITAI described in Sec. \ref{sec_realWorld}, online RL algorithms are used to make a sequence of decisions for each individual over time. The data collected from individuals over time is called \textit{longitudinal data}. 
% I don't think we want to talk about MRTs. In longitudinal data settings, where data are collected longitudinally from each unit in an experiment, \textit{excursion effects} are a common causal estimand used for the primary analysis the adaptively collected data (e.g., data collected using an online RL algorithm) \citep{qian2021estimating,boruvka2018assessing}. 

% \sam{following sentence (I commented it out--but it is in tex file) mixes data from adaptive experiments (e.g. adaptive sampling in bandit environments) with data from the deployment of an adaptive intervention. We need to be really clear in our language.  }  
%While there has been significant work on inference after adaptive sampling for simpler environments such as bandit environments, 
There has been relatively little work on how to conduct inference on data resulting from deployment of an adaptive intervention that included a pooling RL algorithm (i.e., algorithm pools data online over individuals to learn).  
Such data resemble longitudinal datasets, in that each individual experiences a sequence of actions over time and observations are independent across individuals at baseline. However, the resulting data do not follow a classical longitudinal structure.
%The data looks like alongitudinal dataset in which each individual experiences  a sequence of actions and the data from different individuals are independent. However, this is not classical longitudinal data.  
Instead, the pooling mechanism used by the RL algorithm induces between-individual dependency in the longitudinal data. 
As noted in Sec.~\ref{sec:bias-variance}, a pooling RL algorithm selects actions for each individual by leveraging the aggregated history of all individuals at each time point (see Fig.~\ref{fig:pooling}). 
%Pooling data across heterogeneous individuals can smooth out individual-level noise and improve online learning efficiency via the bias–variance tradeoff (i.e., incur bias for each individual due to mismatch between group average and individual -- but reduce variance for sample-efficient learning). 
%\textbf{Statistical inference challenges after pooling RL.} 
While such pooling can substantially improve learning (by trading bias versus variance), it poses significant challenges for statistical analysis: Even when participants or units in the deployment are independently sampled from a population of interest, the assigned interventions or actions become dependent due to the dependence of the policies at each time point on the shared history of all users, as illustrated in Fig. \ref{fig:pooling}. This induces algorithm-driven dependencies across user trajectories over time, which are particularly difficult to characterize---especially when the model used by the RL algorithm is not well-specified, as is often the case in real-world applications. As a result, classical inference methods that assume independent user trajectories are generally invalid.

This is an exciting new but thorny direction for statistical inference methods  \citep{zhang2022statistical, zhang2025replicable}.
\citep{zhang2022statistical, zhang2025replicable}
restricted their work to simple pooling RL algorithms that trade bias with variance  to learn. In their case the pooling algorithm was an RL algorithm using a discount rate, $\gamma=0$ with a misspecified reward model).  They show that if the policy parameterization is smooth, such as  Boltzmann sampling (or softmax exploration \citep{sutton2018reinforcement}) then under regularity conditions, statistical inference (i.e., consistency and asymptotic normality) for marginal parameters based on Z-estimators   holds.  
%Again  the pooling  RL algorithm is using a misspecified model to learn. %\citep{zhang2022statistical, zhang2024replicable}. 
However, the asymptotic variance is inflated relative to the case of independent users due to dependencies introduced by  pooling.

Despite the above findings, several key questions remain. In general, it is an open challenge to develop statistical inference methods that are powerful, robust to model misspecification, and can be applied to data after deployment of any of a large class of RL algorithms. The methods presented by \citet{zhang2022statistical, zhang2025replicable} apply to a specific class of algorithms that form policies that are smooth in their parameters. Additionally, a more detailed analysis is needed to assess the replicability and inferential validity of many practical RL algorithms, such as RL algorithms that use mixed effects to balance pooling and personalization across individuals \citep{tomkins2021intelligentpooling, ghosh2024rebandit}. 
A final open question we note is the need to investigate how reward maximization should be effectively balanced with replicability (discussed under Misspecification in Section~\ref{sec:bias-variance}) in the practical deployment of pooling RL algorithms.

% \sam{not sure if following is about pooling or would concern deployments of individual specific RL algorithms.  Need to be clear about this.} 
% A follow-up question involves statistical power: once a replicable RL algorithm has been deployed, how can statistical power be improved for after-study inference, especially under realistic constraints? A final open question we note is the need to investigate how reward maximization should be effectively balanced with replicability in the practical deployment of RL algorithms. \sam{above is covered in Section~\ref{sec_sequential_design} ....  Should this text or some of the commented out text go there?  OR should that text go here?}

\subsection{Leveraging offline data for algorithm selection and initialization}
\label{sec::algSelect}
Under challenge \textbf{C1} where extensive interaction with the environment is infeasible, offline data from prior deployments provide an important resource for improving early-stage performance in new deployments (Fig. \ref{fig_intro}).
%In the essence of Challenge \textbf{(1)} on the inability to interact massively with the environment, we want to improve the few sample performance of a new deployment using the existing datasets. 
Broadly speaking, there are two ways in which offline data can be leveraged: %to improve the efficiency: 
(1) We can use offline data to help with online algorithm selection and hyper-parameter tuning; (2) We can use the  offline data to warm-start an RL algorithm. 

% Deploying an online RL algorithms in real-life requires careful algorithm selection and tuning. 
% Generally the selection and tuning of the RL algorithm is completed offline (e.g. choosing hyperparameters, feature representations, or exploration strategies) using prior deployment data, other data sources and domain knowledge. 
% \textbf{Offline algorithm selection.} 
\textbf{Selecting an online algorithm via offline data.} We refer to the use of an offline dataset for the selection and tuning of the RL algorithm (e.g. choosing hyperparameters, feature representations, or exploration strategies) as offline algorithm selection \citep{mandel2016offline}. Formally, offline algorithm selection takes as input an offline dataset and a set of candidate learning algorithms, and produces a ranking of these algorithms by their expected online performance (cumulative reward) if deployed. In practice, it allows practitioners to compare multiple RL approaches before deployment. Here we discuss offline algorithm selection specifically for adaptive interventions.

The small but growing literature on offline algorithm selection has largely approached the problem by \emph{constructing a high-fidelity simulator} from the offline data and then running candidate algorithms in this proxy environment. In practice, multiple simulator variants may be constructed to reflect different assumptions about plausible distribution shift or domain shift between past deployments and future implementations. For example, \cite{wang2022no} proposed a nonparametric $k$-nearest neighbor simulator. Given a state–action pair, the simulator retrieves the $k$ most similar state–action pairs from the offline dataset and samples the next state and reward based on what was observed in offline data. \cite{mandel2016offline} introduced a \emph{queue-based simulator} for offline evaluation in recommendation systems. \cite{tang2022towards} proposed a \emph{rejection sampling simulator}, in which the offline dataset is treated as a source of data trajectories obtained from use of a known behavior policy, and the next state and reward is sampled based on a rejection sampling method to mitigate the mismatch between the behavior policy  and the  algorithm's policy. In applied research focused on deciding which RL algorithm to next deploy, \citet{liao2020personalized,trella2022designing,ghosh2024rebandit,xu2025reinforcement} developed simplified individual-specific mechanistic models that allow domain knowledge to be incorporated in the simulator. These works have been used to inform the design of RL agents that have been or will be deployed in real studies.

Simple linear and Markovian dynamic modeling can be unrealistic.
Many simulator-based and offline-to-online approaches share a common limitation: they generate next states and rewards based only on the \emph{current} state–action pair, effectively assuming the environment is stationary and Markovian. As a result, they cannot reproduce nonstationary dynamics or non-Markovian structures, which may be important for realistic algorithm performance. Other simulators use highly nonparametric models, which can be impractical in real-world settings with only small amount of offline data. A promising future direction is to use pretrained deep models that better capture complex system dynamics while leveraging simple  models learned from small amounts of offline data to provide a warm start. There is a growing body of literature on the potential use of open-sourced pretrained foundation models in various data-scare domains like healthcare for prediction \citep{xu2024relcon} and for constructing electronic health records simulators \citep{huang2025evaluation}. Many vision-language models (VLM) have been developed to facilitate robotic control problems \citep{brohan2022rt,nair2022r3m,zitkovich2023rt,kim2024openvla}. A general review can be found in \citep{he2024foundation}. Some use general purposed large language models, or their fine-tuned version on domain specific datasets, which connect to the discussion in Sec. \ref{sec_LLM}. However, much of the existing literature focuses on robotic control tasks with large-scale datasets and relatively accurate simulators derived from physical laws. There is limited work on applying these approaches to directly inform deployed reinforcement learning agents in domains with limited sample sizes (challenge \textbf{C1}).

Extending these approaches further, there is growing interest in designing simulators or models that continually evolve to mirror the evolution of the real-world setting. This is relevant to the discussion in Sec.~\ref{sec_sequential_design} and is related to the concept of \textit{digital twins}. Instantiations of this idea include closed-loop Sim2Real \citep{Chebotar2018} and, for adaptive interventions, the JITAI-Twin framework \citep{Gazi_JITAITwins}. JITAI-Twins are simulation environments repeatedly updated based on data observed in each deployment. Each environment is intended to approximate how a target subpopulation may respond in the next deployment to a JITAI and thus inform decisions on online RL algorithm selection \citep{Gazi_JITAITwins} for the next deployment. %This is relevant to digital twins because data can be used offline to construct an initial virtual representation of the target subpopulation. Candidate online algorithms can then be evaluated in this simulation environment to make decisions for which algorithm is ultimately deployed. 
Upon deployment, newly collected data can be used to update the simulator, and the updated simulator can be used to evaluate design decisions for the following deployment. This creates a bidirectional cycle characteristic of digital twins: real deployments inform the simulator, and the simulator informs future real-world deployments.

% \textbf{Initializing an online algorithm through an offline data.}  Offline data may also be used to warm-up an RL algorithm. This idea appears under several names---hybrid offline–online RL, meta-RL, offline pretraining, and prior-based RL---but all share the goal of improving sample efficiency and stability during online optimization. While these two uses of offline data differ in timing and objective, they rely on similar assumptions about data quality and face closely related challenges.

\textbf{Initializing an online algorithm using offline data.}  Offline data may also be used to warm-start an RL algorithm by enriching existing knowledge.
%to support decision making and exploration during the online learning.
This idea appears under several names---hybrid offline–online RL, meta-RL, offline pretraining, and prior-based RL---but all share the goal of improving sample efficiency during online learning. 
% While these two uses of offline data differ in timing and objective, they rely on similar assumptions about data quality and face closely related challenges. Many of the same modeling choices arise when offline data are used to initialize online RL algorithms. 
A typical approach pretrains policies or value functions on offline data, then fine-tunes them online while constraining updates to remain close to behaviors observed offline; see \citep{nair2020awac, lee2022offline} for deployments of these ideas in robotics and 
\citep{ball2023efficient, nakamoto2023cal} for use on benchmark tasks. Other promising theoretical work, with demonstrations on benchmark tasks, aims to learn transferable representations from offline data \citep{dorfman2020offline, mitchell2021offline, pong2022offline}. Promising Bayesian and prior-based formulations initialize the agent’s belief from offline data and update this belief during online interaction  \citep{hu2024bayesian, kveton2021meta,trella2022designing,ghosh2024rebandit}, 
including approximate Bayesian approaches that tradeoff computational tractability and uncertainty quantification. A majority of above results are only validated in simulated robotic control problems.   \citet{lazic2018data,chen2025offline} presented evidence of benefits from offline-to-online approaches in real industrial  deployments.

There is little systematic understanding of how either offline algorithm selection or offline-to-online  performs under \emph{domain shift}. Domain shift refers to changes in the data-generating process across deployments, such as changes in population characteristics, transition dynamics, or expansions of the action and state space due to new technology advancement \citep{Gazi2025}. 
% In contrast, for offline-to-online RL, domain shift refers to changes between the environment that generated the offline data and the environment encountered during online learning. These shifts present challenges that are distinct from those studied in offline policy evaluation, which focuses on estimating the value of a fixed policy.
Another central challenge shared by the above problems lies in adaptation under partial coverage of the state–action space. When only portions of the state–action space are represented in the offline dataset, offline algorithm selection may favor overly conservative algorithms, while offline-to-online learning may induce overly cautious early behavior. Conversely, aggressive exploration can be unsafe when offline data fail to represent certain regions. Effective methods therefore need to adapt exploration to account for uncertainty induced by offline data limitations.

\textbf{Lack of practical guidelines and benchmarks.} Finally, there is a need for empirical standards and benchmarks that evaluate both offline algorithm selection and offline-to-online learning under conditions likely to occur in health applications—such as sparse or biased offline data, limited online interaction, and nonstationary dynamics. Developing such benchmarks, particularly in data-scarce domains, would accelerate progress toward generalizable algorithms that can safely and effectively leverage existing data to improve online performance in healthcare and online education applications, as discussed in Sec.~\ref{sec_realWorld}.

\section{Iterative Deployment and Continual Improvement}
% \section{Sequential Deployments and Continual Improvement}
\label{sec_sequential_design}
%\sam{we are not doing sequential experimental design.....}
% \ziping{A particular highlight of this selective review is the urgent need for online algorithm design in the context of a sequence of implementations in which the environments may significantly shifts both within and across implementations. These shifts may involve changes in transition dynamics, action spaces, state spaces, or even in the definition of the reward itself (see case discussion in Section \ref{sec_realWorld}). Such sequential implementation settings complicate online agent design in several ways: (i) the agent must continually adapt to the changing environment, which requires both transferring knowledge from previous implementations and acquiring new knowledge in ongoing ones; and (ii) because environments may differ across implementations, there are inherent tradeoffs in objectives between them. Experimental design must explicitly account for these tradeoffs.}

The previous sections discussed (i) online learning within a single deployment, and (ii) offline learning using previously collected data from a prior or ongoing deployment. In this section, we discuss literature that begins to merge these two concepts. We first discuss methods that are designed to tradeoff online learning and post-trial offline learning with data collected during prior deployments. Specifically, algorithms that are optimal for online regret minimization are known to be suboptimal for post-deployment objectives like best-arm identification and estimating treatment effects. Second, we discuss continual learning, which can be thought of as designing algorithms that are able to perform well across distribution shifts and new tasks the algorithm encounters over time. Such algorithms must use previously collected data to inform future decision-making, but realize that previous information may be outdated.

\subsection{Tradeoff Between Optimization and Exploration Objectives of Algorithms}
\label{sec::tradeoffOnlineOffline}
% regret vs policy learning vs continual learning

Within-deployment learning in Sec.~\ref{sec_within_trial} mainly focused on maximizing cumulative reward (i.e., minimizing cumulative regret), while between/post deployment learning methods in Sec. ~\ref{sec_post_trial} focus on statistical inference or offline policy learning.
Interestingly, accomplishing each of these two goals often requires opposing strategies. 
For instance, maximizing cumulative reward emphasizes exploiting current information and exploring only to maximize reward later within the same deployment, whereas learning the optimal policy offline after the deployment is complete calls for pure exploration as the optimal behavior policy.
In this subsection, we discuss the tradeoff between these within- and between-deployment goals.

\textbf{Optimization objectives.}
An important class of tradeoffs in adaptive experimentation arises between cumulative regret minimization and statistical inference for the treatment effect using the adaptively collected experimental data.
% \citet{yao2021power} proposed power-constrained bandits, which personalize interventions online while ensuring sufficient statistical power to detect treatment effects in contextual bandits. 
\citet{yao2021power} demonstrated the tradeoff between cumulative regret minimization during a trial and guaranteeing the power of detecting treatment effect after the trial in a contextual bandit environment.
% , and proved that bounding randomization probabilities is a sufficient approach to ensuring the desired statistical power. 
% and propose meta-algorithms that achieve optimal regret within the class of power-preserving policies.
\citet{simchi2023multi} formulated the tradeoff between regret minimization and the statistical power of detecting the gap between treatment effects of different arms in multi-armed bandit environments as a minimax multi-objective optimization problem, deriving necessary and sufficient conditions for Pareto-optimal solutions. 
\citet{erraqabi2017trading,caria2024adaptive} studied the tradeoff between maximizing cumulative reward and minimizing the estimation error of the average outcome in multi-armed bandits environments.
% \citet{caria2024adaptive} introduced adaptive targeted experiments to improve formal employment outcomes for refugees and local jobseekers by applying multi-armed bandits within strata, and established a tradeoff between participant welfare during the experiment and the asymptotic variance of the average outcome.
% \citet{dai2023clip} proposed an adaptive Neyman allocation to construct the average treatment effect (ATE) estimator, so that the estimator variance converges to the (infeasible) optimal non-adaptive design.
% \citet{erraqabi2017trading} studied the tradeoff between maximizing cumulative reward and minimizing the estimation error across all arms in multi-armed bandits, introducing a convex objective that balances the two goals through a weighting parameter.
Another important tradeoff is between cumulative regret minimization and subsequent offline policy learning. In  experiments, the former reflects the welfare of participants during the trial, while the latter reflects the welfare of the broader population if the policy learned from the adaptive trial is later deployed. Policy learning can be quantified using various metrics, including best-arm identification \citep{qin2024optimizing} and simple regret minimization \citep{gao2022non,athey2022contextual,pacchiano2023experiment,krishnamurthy2023proportional}.
% \citet{qin2024optimizing} investigated the tradeoff between cumulative regret minimization and best-arm identification in multi-armed bandits, showing that by tuning a single scalar parameter—such as in the top-two Thompson sampling algorithm—it is possible to achieve the Pareto frontier between experiment length and total regret.
% \citet{athey2022contextual} designed an adaptive survey experiment on charitable giving using contextual bandits to balance maximizing participant outcomes during the experiment (cumulative regret minimization) and collecting data to learn the best targeting policy for future use (simple regret minimization).
% \citet{gao2022non,athey2022contextual,pacchiano2023experiment} demonstrated the tradeoff between cumulative regret minimization and simple regret minimization in contextual bandits, showing that the balance can be achieved through a tuning parameter that controls the degree of exploration.
% \citet{krishnamurthy2023proportional} further proved that instance-dependent optimal simple regret and minimax optimal cumulative regret cannot be attained simultaneously.
% \citet{pacchiano2023experiment} studied experiment planning, where a sequence of static policies is designed solely from unlabeled context samples without adapting to observed rewards,so that the collected data supports policy learning with near-optimal simple regret. They further showed that two standard adaptive contextual bandit algorithms, which optimize cumulative regret, yield worse simple regret upper bounds compared to the optimal adaptive algorithm for policy learning.

Existing work has mainly considered the tradeoff between cumulative regret and treatment effect inference or policy learning. 
However, other important objectives also require deeper examination, such as off-policy evaluation \citep{uehara2022review}, subgroup identification \citep{spiess2021finding}, and secondary analyses like heterogeneous treatment effect estimation.
A related question involves statistical power: what are realistic constraints on an  an RL algorithm that leads to improved statistical power for after-study inference based on the resulting data?

\textbf{Tradeoffs across a sequence of experiments or MDPs.}
The literature primarily examine tradeoffs between multiple objectives within a single experiment \citep{yao2021power,simchi2023multi,erraqabi2017trading,caria2024adaptive,gao2022non,athey2022contextual,pacchiano2023experiment,krishnamurthy2023proportional,qin2024optimizing}. In contrast, tradeoffs across a sequence of (nonstationary) experiments can be even more restrictive.
\citet{xu2024fallacy} studied the tradeoff between online cumulative regret in one implementation and simple regret (intuitively, the warm-up performance) in a subsequent implementation when the problem setup changes—for example, when new constraints are introduced on the policy set. It remains an open problem whether other forms of continual improvement may impose additional tradeoffs between objectives.

Moreover, the existing literature has largely focused on multi-armed bandit and contextual bandit environments. It is worth exploring how these conclusions extend to other environments such as MDPs or more general settings, either in a single experiment or sequential experiment design.

\subsection{Nonstationarity and Continual Learning} \label{sec_nonstationary_continual}
The literature on nonstationarity and the literature on continual learning both focus on developing \textit{algorithms} that can adapt over time (not just \textit{agents} that can adapt over time via online RL). The nonstationarity learning literature \citep{abbasi2023new,auer2019adaptively, besbes2019optimal,cheung2019learning, trella2024non,chen2023non, jia2023smooth}
generally focuses on developing decision making algorithms under a specific assumption on how the underlying environment may change over time. Rather than focusing primarily on assumptions made about a single environment, continual learning focuses more on the learning algorithm. Continual learning emphasizes that the algorithm must be able to learn to perform well on potentially entirely new tasks encountered over time \citep{abel2023definition}.  Due to the bias versus variance tradeoff as well as computational constraints, a learning algorithm may need to continually adapt over time even in a stationary environment across deployments \citep{kumar2023continual}.
%\sam{the following paragraph is very confusing...unclear what this has to do with continual adaptation in a stationary environment.  I deleted.} Any assumptions can be re-specified in this process. For example, to alleviate the misspecified assumptions in a causal DAG, we can estimate the (partial or full) causal structure with data from prior deployments before learning the policy online \citep{lu2021causal,mendez2023carl} or offline \citep{zhu2022offline,sun2024acamda}, or using causal representation learning or causal state abstraction \citep{zhang2019learning,zhang2021learning,sontakke2021causal,huang2022action,wang2022causal,wang2024building} to compress states. 
%Other lines of work learn the (partial or full) causal structure before learning the policy online \citep{lu2021causal,mendez2023carl} or offline \citet{zhu2022offline,sun2024acamda}.
Continual learning is an emerging area that requires significantly more study. In this section we provide a brief overview of 
%the nonstationary learning literature and 
emerging research on continual RL. 
Continual learning is becoming increasingly relevant when thinking about how machine learning  might be implemented  in  applications ranging from medicine (e.g., \citep{bruno2025continual,lee2020clinical}) to recommender systems \citep{mi2020ader,cai2022reloop}. The desire to study continual learning is motivated by the need for computationally efficient algorithms that are able to continually learn and adapt to new tasks, while retaining skills from previously learned tasks. Work on the theoretical front has emphasized the need to clearly define the explicit computational and memory constraints that motivate the need for continual learning \citep{abel2023definition,kumar2023continual}. In the broader machine learning literature, a concept that is prominent in continual learning is the \textit{stability-plasticity dilemma}. The stability-plasticity dilemma refers to the need for algorithms to balance retaining previously learned knowledge and learning from new information \citep{khetarpal2022towards}. For example, methods have been proposed to manage this dilemma and prevent catastrophic forgetting via experience replay (e.g., \citep{rolnick2019experience,zhang2022simple}) and regularization to encourage only a subset of parameters be updated \citep{ahn2019uncertainty}.  %Multiple open questions remain, particularly in the RL setting. 
%, but several open questions remain. 

%\textbf{Articulating realistic environment structures and the need for realistic simulation environments.} 
The continual learning literature in RL is relatively nascent. As such, there are a variety of significant open problems, of which we discuss a few. See \citet{khetarpal2022towards} for a more comprehensive discussion of open problems. It is unlikely that a single continual learning algorithm will perform well in all possible environments. A significant open challenge is to better articulate the particular problem structures and environments that are a continual learning algorithm is likely to perform well on. In this vein, several researchers advocate for the need to build more realistic simulation environments to capture more realistic dynamics \citep{wolczyk2021continual}---this 
%\sam{rationale for kumar2024need cite about online model selection.}
can improve both algorithm evaluation and algorithm development. Additionally, many 
%nonstationary and
continual learning algorithms have various hyperparameters that significantly impact performance. Developing methods to optimize these hyperparameters or developing algorithms that are more robust to hyperparmeter choices will make these algorithms more deployable in practice. Finally, it is also of interest to develop continual learning algorithms that adapt to, or at least degrade gracefully under potentially misspecified assumptions on the environment.

\section{{Case Study: the HeartSteps JITAI}}
\label{sec:HeartSteps}
In this case study, we describe concretely how many of the statistical challenges and opportunities we present throughout the paper apply to the real-world HeartSteps JITAI introduced in Sec. \ref{sec:adaptInterv}. The overarching goal of the HeartSteps JITAI is to help individuals increase and sustain physical activity in daily life. To do so, HeartSteps uses mobile and wearable technology to deliver walking suggestions and other intervention content to promote physical activity. HeartSteps provides a concrete example of the two challenges introduced in Sec. \ref{sec_intro} and emphasized throughout the paper: \textbf{C1} -- limited interaction with the target environment and \textbf{C2} -- the need for continual redesign and redeployment. HeartSteps cannot explore arbitrarily: intervening too often can burden participants, lead to habituation, and cause disengagement. Furthermore, the target subpopulation, technology and intervention content have  changed across the four HeartSteps deployments.

The sequential decision-making problem is formulated as follows. At each decision time $t$, the HeartSteps JITAI observes information about the individual’s current state and then decides whether to deliver an intervention option such as a walking suggestion. The action, $A_t$, may encode whether an walking suggestion is delivered, the state, $S_t$, may include time of day, location, recent activity, weather, or engagement-related variables as informed by a causal DAG (Fig. \ref{fig:dag}), and the reward, $R_t$, may be a function of subsequent step count, affective valuations of physical activity, or engagement with the HeartSteps JITAI application.

\subsection{{Past Deployments and Lessons Learned}}

The first HeartSteps deployment (``HeartSteps V1" in this paper) illustrates the role of offline analyses prior to an online RL algorithm being deployed (Fig. \ref{fig_intro}). HeartSteps V1 was a clinical trial in which the behavior policy involved pure exploration, rather than the immediate deployment of an online RL algorithm \citep{Klasnja2019}. This is an important note given the higher stakes nature of health-related interventions. Participants were ``micro-randomized" to receive either some type of activity suggestion ($A_t = 1$) or no suggestion ($A_t = 0$) with 60-40 probability at each decision time. Information collected from participants that could potentially later be used as state information in offline analyses ($S_t$) included time of day, day of the week, location, recent physical activity, weather, etc. The short-term outcome following each decision time that was used for offline analyses (i.e., analogous to reward, $R_t$) was the 30-minute step count following each decision time. The resulting sequentially randomized longitudinal data was used to estimate causal effects of the activity suggestions on physical activity and examine how these effects changed over time \citep{Klasnja2019}.

The second HeartSteps deployment (``HeartSteps V2") illustrates the value of knowledge transfer from offline learning to online RL. In HeartSteps V2, the goal was for an online RL algorithm to decide at each decision time whether or not to send a contextually tailored walking suggestion to promote physical activity \citep{liao2020personalized}. The deployment focused on individuals with blood pressure in the stage 1 hypertension range, a population that differed from the healthy sedentary adults in HeartSteps V1. This change in target population is a concrete example of challenge \textbf{C2} introduced in Sec. \ref{sec_intro}: data from the previous deployment were useful, but could not be assumed to exactly represent the new deployment environment.

The HeartSteps V2 online RL algorithm made several statistical compromises that exemplify the bias-variance tradeoffs discussed in Sec. \ref{sec:bias-variance} and are concrete examples of how challenge \textbf{C1} introduced in Sec. \ref{sec_intro} takes effect: extensive exploration is not possible with each recipient of a JITAI, and thus, learning sophisticated models may not be feasible within the specified intervention time frame. For example, using a prediction horizon longer than one timestep could in principle help account for effects of suggestions on future activity, engagement, and burden. However, learning such a MDP-based model or $Q$ function online from a small number of noisy observations per participant would tend towards high variance. Instead a  single MDP model was constructed based on all individuals' data in HeartSteps V1 and the optimal $Q$ function for this  MDP estimated. The deployed algorithm onboard HeartSteps V2/V3  was a person-specific online contextual bandit algorithm  with reward equal to the sum of the log of subsequent 30-minute step count and the estimated optimal $Q$ function from the HeartSteps V1 analysis \citep{liao2020personalized}.  The simplification likely introduced bias due to between population heterogeneity (change in population)  and  within population heterogeneity in optimal Q-functions but reduced variance and improved the feasibility of online learning during during deployment. Furthermore, the algorithm  used information from HeartSteps V1 to select features, construct priors, and build simulation environments for evaluating candidate algorithmic choices before deployment. Concretely, the between-deployment statistical inference following HeartSteps V1 informed the next deployment’s online RL algorithm.

In line with the discussion in Sec. \ref{sec::tradeoffOnlineOffline}, the goal of HeartSteps V2/V3 was to both benefit the current recipients of the JITAI as well as enable causal inference and  offline, off-policy learning in preparation for the next HeartSteps deployment. To do so, the online RL algorithm for HeartSteps V2/V3 selected actions stochastically and bounded action-selection probabilities away from 0 and 1. These design choices likely reduced short-term reward relative to a more exploitative algorithm, but the resulting data could then be more useful for offline, off-policy learning. As discussed in Sec.~\ref{sec::tradeoffOnlineOffline}, there is a tradeoff between cumulative regret minimization and subsequent offline policy learning. Such analyses support online-to-offline (and then -to-online learning): the deployment generates data that can be used for offline causal inference to generate knowledge that can then be used to refine the online RL algorithm for the next deployment.

\subsection{{Opportunities for Ongoing and Future Deployments}}
Causal reasoning, as discussed in Sec. \ref{sec:bias-variance}, is playing an important role in preparation for a fifth deployment of HeartSteps. For example, Fig. \ref{fig:dag} is an example of an early  causal DAG for HeartSteps. The causal DAG is useful in several ways. First, it clarifies which variables should be included in the state representation. Second, it pinpoints short-term outcomes that are plausible mediators for longer-term outcomes that may be too sparse or delayed for online learning. Third, causal DAGs can be used to inform simulations for algorithm selection or sensitivity analyses to understand the impact of specific causal assumptions on algorithm performance. These example use cases have been investigated in recent work motivated by preparation for the fifth deployment  \citep{gao2025harnessing}.
The iterative development of HeartSteps also motivates offline simulation frameworks for online algorithm selection that benefit from ideas in the digital twin literature, as discussed in Sec. \ref{sec::algSelect}. The JITAI-Twin framework, (see Sec. \ref{sec::algSelect}), was motivated by the iterative development of HeartSteps \citep{Gazi_JITAITwins}.
%JITAI-Twins represent simulation environments repeatedly updated based on data observed in each deployment. Each environment is intended to approximate how a target subpopulation may respond in the next deployment to a JITAI and thus inform decisions on online RL algorithm selection \citep{Gazi_JITAITwins} for this next deployment. %This is relevant to digital twins because data can be used offline to construct an initial virtual representation of the target subpopulation. Candidate online algorithms can then be evaluated in this simulation environment to make decisions for which algorithm is ultimately deployed. 
%Upon deployment, newly collected data can be used to update the simulator, and the updated simulator can be used to evaluate design decisions for the following deployment. This creates a bidirectional cycle characteristic of digital twins: real deployments inform the simulator, and the simulator informs future real-world deployments.}

The HeartSteps example also makes clear that continual improvement is not simply repeated online fine-tuning to arrive at some perpetually optimal policy. As discussed in Sec. \ref{sec_nonstationary_continual}, the decision problem itself may change across deployments. The scientific team has removed ineffective intervention options (i.e., changes to the action space), changed the desired outcomes (i.e., modifications of the reward, as in the latest versions of HeartSteps \citep{gao2025active}), incorporated new sensors (i.e., changes to the state space), and altered the deployment target subpopulation (i.e., changes to the underlying environment). These changes require human oversight and offline analyses before subsequent deployments of an online RL algorithm (i.e,. ``between" deployments). The opportunities for use-inspired continual learning research in these settings are exciting. 

\subsection{{Summary}}
The continual improvement of RL algorithms deployed onboard the HeartSteps JITAI illustrates the core message of this paper: the deployment of real-world RL algorithms for interventions that interact with humans is not a one-time exercise in determining some optimal policy offline or converging upon an optimal policy online. It is a statistical learning process that unfolds across deployments of RL algorithms, combining online learning, offline analyses, and human-guided redesign, as depicted in Fig. \ref{fig_intro}. Progress in these settings will require methods that explicitly address this need for continual improvement, taking into consideration the challenges associated with limited data on new individuals, heterogeneity across individuals, and incomplete models of human behavior. All the while, algorithms must preserve the utility of the data generated by each deployment for offline analyses to improve models of human behavior and allow for off-policy learning for offline-to-online-to-offline-to... RL.

\section{Conclusion}
In this survey and perspective, we describe an application-grounded framing of RL that is driven by the realities of deploying RL in data sparse settings and in systems that interact with humans. We draw upon real-world examples to illustrate two central challenges in these settings: \textbf{(C1)} extensive exploration data required for the RL system  to learn solely on each individual and solely online is often unavailable, and \textbf{(C2)} significant changes in the deployment environment may necessitate redesign and redeployment of RL algorithms as a system is continually improved. We accordingly framed practice-oriented RL as a three-component process centered around learning and continual improvement: within-deployment online learning, between-deployment offline analysis, and iterative deployment–redeployment for continual improvement (Fig. \ref{fig_intro}). Organizing relevant work around this framework, we highlighted use-inspired directions that have the potential to impact real-world applications. Taken together, these lines of work point toward a broader research agenda in RL that goes beyond classical regret guarantees to prioritize continual learning, replicability, valid inference, and scientific interpretability. Realizing the potential of RL in high-stakes real-world applications will require closer integration of statistical methodology, domain knowledge, and algorithm design. Our hope is that the presented framework and the highlighted open problems help guide such interdisciplinary efforts.

%% Acknowledgments
\section*{Acknowledgments}

Authors marked with $^\ast$ co-led this work and are listed alphabetically. All other authors are listed alphabetically, except S.\ A.\ Murphy, who is listed last as the corresponding author. S.\ A.\ Murphy holds concurrent appointments at Harvard University and as an Amazon Scholar. This paper describes work performed at Harvard University and is not associated with Amazon.

%% Funding
\paragraph{Funding.} S. A. Murphy's research group is supported by the U.S. National Institutes of Health (NIH) under grants R01HL125440, P50DA054039, P41EB028242, and UH3DE028723. A. H. Gazi is additionally supported by an NIH Award K99EB037411. Y. Guo is supported by the U.S. National Science Foundation (NSF) under Grant DMS-2515285.

%% Bibliography
\bibliographystyle{plainnat}
\bibliography{references}

\end{document}